\documentclass[12pt]{article}

\usepackage[english]{babel}
\usepackage{amsmath,amsthm}
\usepackage{amsfonts}
\usepackage{graphicx}
\usepackage{amssymb}

\newcommand{\dsum}{\displaystyle\sum}
\newcommand{\dint}{\displaystyle\int}

\begin{document}

\title{\textbf{Explicit force formlulas for two dimensional potential flow with
multiple bodies and multiple free vortices}}
\author{\textbf{Chen-Yuan BAI, Juan Li and Zi-Niu WU\thanks{ziniuwu@tsinghua.edu.cn.}}\\
School of Aerospace, Tsinghua University, Beijing, 100084, PR China} 
\renewcommand{\today}{April 19, 2013}

\maketitle

\begin{abstract}
For problems with multiple bodies, the current integral approach needs the
use of auxiliary potential functions in order to have an individual force
formula for each body. While the singularity approach, based on an extension
of the unsteady Lagally theorem, is restricted to multibody and multivortex
flows without bound vortex and vortex production. In this paper, we\
consider multibody and multivortex flow and derive force formulas, in both
forms of singularity approach and integral approach but without {\normalsize %
auxiliary function}, that give individual forces of each body for unsteady
two dimensional potential flow with vortex production on the surface of
bodies. A number of problems, including Karman vortex street, Wagner problem
of impulsively starting flow, interaction of two circular cylinders with
circulation, and interaction of an airfoil with a bound vortex, are used to
validate the force formulas.
\end{abstract}


\section{Introduction}

In the classic Kutta Joukowski theorem, the role of the starting vortex,
produced during the starting up of flow, is omitted by simply assuming it
disappear in the far flow field. Thus the lift is related to the circulation
of bound vortex ({\normalsize Batchelor 1967). Under the assumption of
steady potential flow, the circulation of the bound vortex is determined by
the Kutta condition, due to which the role of viscosity is implicitly
incorporated though explicitly ignored (Grighton 1985). The lift predicted
by Kutta Joukowski theorem within the framework of inviscid flow theory is
quite accurate even for real viscous flow, provided the flow is steady and
unseparated, see Anderson }(1984,p.192) for more details.

For many problems, there may be free vortices, including the starting
vortex, or other bodies close to the body. These problems have been
attracting great attentions since more than two decades ago, due to their
wide applications in unsteady flows (Chow\&Huang 1982, Lee\&Smith 1991, Aref
2007) and in multibody flows such as multi-turbine flow (Oterberg, 2010),
multi-blade flow (Smith \& Timoshin, 1996), \ multi-element airfoil flow
(Katz \& Plotkin, 2001), multi-wing aerodynamics as for dragonfly (Hsieh,
Kung\&Chang 2010), and flows in staggered cylinders (Crowdy 2006).
{\normalsize \ Early studies considering the interaction of a vortex with a
wing can be found in Saffman (1992,p122). The recent studies led to force
formulas which can be conveniently classified into integral approaches and
singularity approaches. For integral approaches, the forces are expressed in
terms of the time variation of the integrated vorticity moment or fluid
impulse. For singularity approaches, the forces are expressed in terms of
the production of the strengths of singularities and the speeds of the real
and image singularities. In both approaches, there may be additional terms,
including for instance the added mass effect, if the body is subjected to
acceleration, rotation and deformation. }

{\normalsize \ The major differences between various integral approaches lie
on the choice of the integration domain. The integral approaches of Wu
(1981) for viscous flow and Saffman (1992) for inviscid flow use integrals
defined for the whole space including both the fluid region and the solid
region. With the help of auxiliary flow potentials, Howe (1995) derived an
integral approach with a surface integral for the viscous force but with the
volume integral only defined for the space occupied by the fluid and with
the contribution from added mass, pressure by vortices and skin frictions
represented by separate terms. Such an approach has been subsequently
extended to multiple bodies by Ragazzo \& Tabak (2007) and Chang, Yang \&
Chu (2008). {\small \ }The advantage of the integral approach is that it
only requires the knowledge of the velocity field and its derivatives. The
disadvantage is that the vortices in the far flow field, such as starting
vortices, must also be taken into account, even when the starting vortices
are at infinity. Integral approaches for a truncated domain are then
derived, at the sacrifice of introducing boundary surface integrals (}Noca
et al 1999,{\normalsize {\small \ }Wu et al 2007, Eames et al 2008), see Wu,
Lu\& Zhuang (2007) for more recent advances related to integral approaches
and for a thorough discussion of the usefulness of the various integral
approaches.}

The force formulas by the singularity approach are basically worked out
through using the complex potential theory and the unsteady Blasius theorem.
The Blasius equation for the general case of unsteady flow and for a body in
arbitrary motion can be found in Thomson (1968). Streitlien and
Triantafyllou (1995) derived a force formula for a single Joukowski airfoil
surrounded with point vortices convected freely. Ramodanov (2002) considered
the motion of a circular cylinder in the presence of N point vortices and
the forces are expressed in terms of the speeds of both real and image
vortices. {\normalsize Kanso\&Oskouei (2008) extended these formulas to
deformable bodies with vortex production. The forces are also expressed in
terms of the velocities of real vortices plus the time variation of an
additional integral term representing all the effects other than the motion
of vortices outside of the cylinder, see Shashikanth et al (2002) for a
circular cylinder, Borisov et al (2007)} {\normalsize \ for a general
cylinder, and Michelin\& Smith (2010) for problems with vortex production. }

For singularity approaches, both the real and image vortices are implicitly
or explicitly included locally according to their real positions. For the
case of an airfoil interacting with one outside vortex, Katz and Plotkin
(2001, chapter 6.9) express the force in terms of the induced velocity at
the body center.\ The role of real vortex outside of the body is represented
by the induced velocity at the body center. This result was obtained under
the lumped vortex assumption and extended to the case of multi-airfoil and
multi-vortex flow by Bai\&Wu (2013). The presence of singularities such as
sources and doublets outside of a body has been also studied in the
framework of Lagally theorem (Milne-Thomson 1968, Landweber\&Miloh1980). Wu,
Yang \&Young (2012) extended the Lagally theorem to the case of two
dimensional flow with multibody moving in a still fluid in the presence of
multiple free vortices, and the forces are expressed in terms of the induced
velocities or its derivatives at the positions of the internal
singularities, including sources, doublets and image vortices. Bound vortex
and vortex production are not considered in this work. Moreover, validation
and application studies are restricted to circular cylinders.

For convenience, the approaches purely based on the velocities of
singularities will be called \emph{singularity velocity method}. When the
induced flow velocities at the inner singularities are used to replace the
velocities of the free singularities, the approach will be called \emph{%
induced velocity method}.

{\normalsize In this paper, we consider unsteady two dimensional potential
flow with multiple bodies and multiple free vortices. Each body is assumed
to have an arbitrary shape and vortex production is also considered. Force
formulas, algebraic and explicit for each body, valid for both discrete and
continuously distributed singularities, will be derived. This paper will be
organized as follows.}

In section 2 {\normalsize \ we first use a momentum approach to relate the
lift force and induced drag force to the speeds of singularities inside and
outside of a single body (singularity velocity method) with and without
vortex production. }We then relate the force terms due to the outside
singularities to the induced velocities inside the body and express the
forces in terms of the relative induced velocities and strengths of
singularities inside of the body (induced velocity method). The {\normalsize %
induced velocity method} is then extended to the case of multibody flow.
Both the singularity velocity method and the induced velocity method, for
discrete singularities, are finally extended to the case of continuously
distributed vortices, sources and doublets. {\normalsize \ }

In section 3, we use three problems to show how to use or to validate the
force formulas. The first problem is for circular cylinders for which the
method of images can be used to obtain the singularities. Notably, we will
consider the interaction of two circular cylinders with given bound
vortices, for which Crowdy (2006) gives exact solution. The second problem
is the drag for Karman vortex street, a very difficulty problem since the
drag is not related to the shape of the body. \ The last problem is the
interaction of a free vortex with an airfoil, including the well known
example of impulsively starting flow with vortex shedding. Finally we will
consider an application, a bound vortex above the middle point of a flat
plate.

A short summary, with emphasis on the new features of the present work and
remaining work to be done, is provided in section 4.

\section{Force formulas in various forms}

{\normalsize In this section, we first state the flow field to be
considered. Then we use a momentum approach based on a suitably designed
control volume to obtain a force formula. Then this force formula is
rewritten in a form such that the forces are only related to flow properties
inside the body. Finally the force formulas will be extended to multibody
flow and to flows with continuous distribution of vortices, sources and
doublets. Important remarks, including relation and difference to other
theories and the treatment of vortex production forces, will also be
provided. }

\subsection{Description of the flow field}

Consider {\normalsize a body with infinite span, immersed in an
incompressible two-dimensional flow } {\normalsize at} constant density $%
\rho $. The{\normalsize \ freestream velocity $V_{\infty }$ is assumed
horizontal. The local flow field is supposed to be generated by vortices,
sources, doublets and body acceleration and rotation, in a way that the
total velocity of the flow can be obtained by a linear superposition of the
induced velocities due to these factors. Singularities, including point
vortices, sources and doublets, are assumed to be either inside of the body
(called inner ones) or outside of the body (called outer ones). Each of the
outer singularities will be assumed to be at a finite distance to the body. }

{\normalsize The sum of the strengths of the inner vortices is equal to}
{\normalsize $\Gamma _{b}=\int_{\partial A}\left( udx+vdy\right) $, which is
just the circulation of the bound vortex, the closed curve $\partial A$ is
along the body with an anticlockwise path, so that a clockwise circulation
has a negative sign.\ }We note that even when there is vortex production,
the conservation of total circulation holds
\begin{equation}
\underset{i}{\sum }\frac{d\Gamma _{i}}{dt}=0.  \label{eq-gamma}
\end{equation}

A singularity, located at $(x_{i},y_{i})$ but generally moving at the
velocity $\left( \frac{dx_{i}}{dt},\frac{dy_{i}}{dt}\right) $, will be
either a point vortex of strength $\Gamma _{i}$, a source of strength $m_{i}$%
, or a doublet of strength $\mu _{i}$.

The (fluid) velocity induced at $(x,y)$ by a point vortex $(i)$ at $%
(x_{i},y_{i})$ is%
\begin{equation}
\left\{
\begin{array}{c}
u^{(i)}(x,y)=\frac{\partial \psi _{i}}{\partial y}=\frac{\partial \phi _{i}}{%
\partial x}=-\frac{\Gamma _{i}}{2\pi }\frac{y-y_{i}}{r_{i}^{2}} \\
v^{(i)}(x,y)=-\frac{\partial \psi _{i}}{\partial x}=\frac{\partial \phi _{i}%
}{\partial y}=\frac{\Gamma _{i}}{2\pi }\frac{x-x_{i}}{r_{i}^{2}}%
\end{array}%
\right.  \label{eq-flow-v}
\end{equation}%
\ where $\psi _{i}=-\frac{\Gamma _{i}}{2\pi }\ln r_{i}$ with $r_{i}=\sqrt{%
(x-x_{i})^{2}+(y-y_{i})^{2}}$ is the stream function and $\phi _{i}=\frac{%
\Gamma _{i}\theta _{i}}{2\pi }$ is the velocity potential. The angle $\theta
_{i}$ is defined such that $x-x_{i}=r_{i}\cos \theta _{i}$ and $%
y-y_{i}=r_{i}\sin \theta _{i}$. \ It should be emphasized that this induced
velocity is independent of the velocity of the vortex.

A doublet can be treated equivalently as a vortex pair ({\small {\normalsize %
Thomson 1968,p361), see at the end of section 2.3. Thus we first assume the
doublets have been transformed into vortices and just derive forces due to
vortices and sources, then the explicit influence due to doublets will be
derived directly from the vortex based forces (section 2.3). }}

Pure source (sink) singularities are in fact not required since we only
consider closed bodies, for which we may always use a number of source
doublets to represent pure sources. But for completeness we will also
consider the existence of sources. The flow field due to a point source $(i)$
of strength $m_{i}$ is
\begin{equation}
\left\{
\begin{array}{c}
u^{(i)}(x,y)=\frac{\partial \psi _{i}}{\partial y}=\frac{\partial \phi _{i}}{%
\partial x}=\frac{m_{i}}{2\pi }\frac{x-x_{i}}{r_{i}^{2}} \\
v^{(i)}(x,y)=-\frac{\partial \psi _{i}}{\partial x}=\frac{\partial \phi _{i}%
}{\partial y}=\frac{m_{i}}{2\pi }\frac{y-y_{i}}{r_{i}^{2}}%
\end{array}%
\right.  \label{eq-flow-s}
\end{equation}%
Since the functional forms for (\ref{eq-flow-s}) and (\ref{eq-flow-v}) are
similar, the forces due to sources can be similarly obtained as for vortices.

Now consider body generated flow, for a body ($A$) rotating at the angular
speed $\Omega $ (positive if anticlockwise) around the point ($x_{o},y_{o}$)
which translates in addition at velocity $(U,V)$ in a flow already with a
free stream velocity $V_{\infty }$. \ The flow potential and stream function
due to body translation and rotation may be decomposed as

\begin{equation}
\phi _{b}=U\phi _{U}+V\phi _{V}+\Omega \phi _{\Omega }\text{, \ }\psi
_{b}=U\psi _{U}+V\psi _{V}+\Omega \psi _{\Omega }\text{\ }
\label{eq-ps-body}
\end{equation}%
where $\phi _{U},\phi _{V},\phi _{\Omega }$ and $\psi _{U},\psi _{V},\psi
_{\Omega }$ are the so called normalized \ potentials and stream functions,
generated by the body translating and rotating at unitary speed. The forces
due to this will be related to added mass effects.%
\begin{figure}[ptb]
\centering
\includegraphics[width=0.6\textwidth]{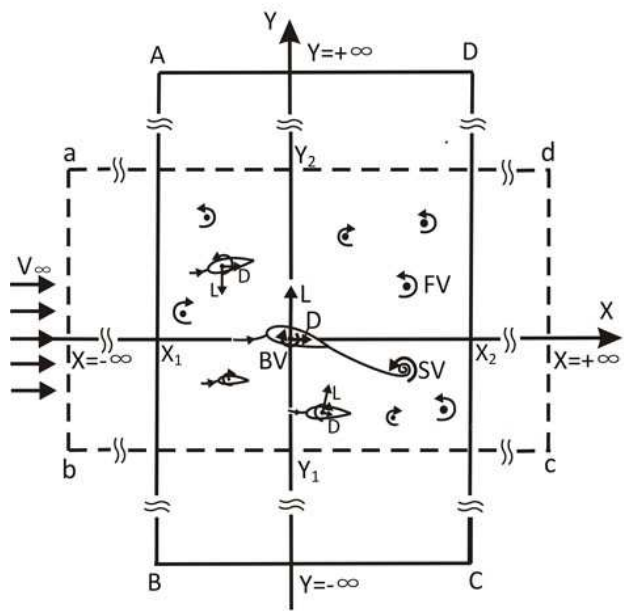}%
\caption{C{\protect\small control volumes:
the solid box ABCD defines a vertical control volume }$(X_{1},X_{2})\times
(-\infty ,\infty )${\protect\small \ \ and the dashed box abcd defines a
horizontal control volume }$(-\infty ,\infty )\times (Y_{1},Y_{2})$.}%
\label{fig1}%
\end{figure}%

\subsection{Singularity velocity method}

Now we use a momentum balance approach, based on the control volumes
(vertical control volume and horizontal control volume) defined in Fig.1, to
study the forces. The use of two types of control volumes is just for
simplification of some algebraic operations during the derivation of forces.

We assume the boundaries of the control volume to be far enough away from
the body and singularities, so that the momentum balance approach used here
will be linear and therefore the contributions by various singularities
(when the body is regarded fixed) and by body generated flow (when the body
is accelerating and rotating) can be decomposed.

\emph{A) forces due to vortices}. For lift, we use the vertical control
volume for momentum balance. The body is subjected to a lift force $(L_{v})$%
, due to vortices, so that the fluid in the control volume is subjected to a
force of equal magnitude but with an opposite direction, and this force is
balanced by the momentum flux across the left and right boundaries $(x=X_{1}$%
, $x=X_{2}$) and the time variation of the momentum inside the control
volume excluding the body, i.e.,

\begin{equation}
{\normalsize L}_{v}{\normalsize =\rho \underset{i}{\sum }\int_{x=x_{1}}V_{%
\infty }v^{(i)}dy-\rho \underset{i}{\sum }\int_{x=x_{2}}V_{\infty }v^{(i)}dy-%
}\underset{i}{\sum }\frac{dm_{y}^{(i)}}{dt}+L_{av}  \label{eq-img-0}
\end{equation}%
The last two terms on the right hand side represent the momentum change in
the control volume excluding the region occupied by the body. Hence if $%
\frac{dm_{y}^{(i)}}{dt}$ is defined for the whole space in the control
volume, i.e.,
\begin{equation*}
\frac{dm_{y}^{(i)}}{dt}=\frac{d}{dt}\overset{X_{2}}{\underset{X_{1}}{\int }}%
\overset{\infty }{\underset{-\infty }{\int }}\rho v^{(i)}dydx
\end{equation*}%
then, $L_{av}$ represents the momentum variation rate of the fictitious
fluid inside the body. The specific role of $L_{av}$ will be further
discussed in the end of this subsection.

To find the explicit form of the integrals involved in (\ref{eq-img-0}), we
use the identity{\small {\normalsize
\begin{equation*}
\int_{-\infty }^{\infty }\frac{c}{b^{2}+(y-d)^{2}}dy=\frac{\pi c}{\left\vert
b\right\vert }
\end{equation*}%
which holds for any set of parameters }} $b,c,d$ independent of $y$. Hence%
\begin{equation*}
\left\{
\begin{array}{l}
\rho \overset{\infty }{\underset{-\infty }{\int }}\left. V_{\infty
}v^{(i)}\right\vert _{X_{1}}dy=\frac{\rho V_{\infty }\Gamma _{i}}{2}\frac{%
X_{1}-x_{i}}{\left\vert X_{1}-x_{i}\right\vert }=-\frac{1}{2}\rho V_{\infty
}\Gamma _{i} \\
\rho \overset{\infty }{\underset{-\infty }{\int }}\left. V_{\infty
}v^{(i)}\right\vert _{X_{2}}dy=\frac{\rho V_{\infty }\Gamma _{i}}{2}\frac{%
X_{2}-x_{i}}{\left\vert X_{2}-x_{i}\right\vert }=\frac{1}{2}\rho V_{\infty
}\Gamma _{i} \\
\frac{d}{dt}\overset{X_{2}}{\underset{X_{1}}{\int }}\overset{\infty }{%
\underset{-\infty }{\int }}\rho v^{(i)}dydx=\rho \frac{1}{2\pi }\frac{d}{dt}%
\left( \Gamma _{i}\overset{x_{i}}{\underset{X_{1}}{\int }}\frac{\pi (x-x_{i})%
}{\left\vert x-x_{i}\right\vert }+\Gamma _{i}\overset{X_{2}}{\underset{x_{i}}%
{\int }}\frac{\pi (x-x_{i})}{\left\vert x-x_{i}\right\vert }\right) dx \\
\text{ \ \ \ \ \ \ \ \ \ \ \ \ \ \ \ }=\rho \frac{1}{2\pi }\frac{d}{dt}%
\left( -\pi \left( x_{i}-X_{1}\right) \Gamma _{i}+\pi \left(
X_{2}-x_{i}\right) \Gamma _{i}\right) \\
\text{ \ \ \ \ \ \ \ \ \ \ \ \ \ \ \ }=-\rho \Gamma _{i}\frac{d\left( \Gamma
_{i}x_{i}\right) }{dt}+\rho \frac{1}{2\pi }\left( \pi (X_{2}+X_{1})\right)
\frac{d\Gamma _{i}}{dt}%
\end{array}%
\right.
\end{equation*}%
{\small {\normalsize Inserting these formulas into (\ref{eq-img-0})\ we get}}%
\begin{equation*}
L_{v}=-\rho \underset{i}{\sum }\left( V_{\infty }\Gamma _{i}-\frac{d\left(
\Gamma _{i}x_{i}\right) }{dt}\right) +L_{av}+\rho \frac{1}{2\pi }\left( \pi
(X_{2}+X_{1})\right) \underset{i}{\sum }\frac{d\Gamma _{i}}{dt}
\end{equation*}%
{\small {\normalsize Using (\ref{eq-gamma}) to eliminate the last term on
the right hand side we obtain }}%
\begin{equation}
L_{v}=-\rho \underset{i}{\sum }\left( V_{\infty }\Gamma _{i}-\frac{d\left(
\Gamma _{i}x_{i}\right) }{dt}\right) +L_{av}  \label{eq-force-lift-v}
\end{equation}%
Similarly, by using the horizontal control volume, we obtain the drag force
formula

{\normalsize
\begin{equation}
D_{v}=-\rho \underset{i}{\sum }\frac{d\left( \Gamma _{i}y_{i}\right) }{dt}%
+D_{av}  \label{eq-force-drag-v}
\end{equation}%
where }$D_{av}$ represents the change of $x$-momentum inside the body.

B) \emph{forces due to sources. \ }As remarked in the last subsection, the
functional form of the velocity components $u$ and $v$ due to a point source
is the same as that of $v$ and $-u$ for a\ vortex. Hence we can use the
results of vortices and directly write down the force formulas for point
sources as

\begin{equation}
L_{st}=\rho \underset{i}{\sum }\frac{d\left( m_{i}y_{i}\right) }{dt}%
+L_{as},D_{st}=\rho \underset{i}{\sum }\frac{d\left( m_{i}x_{i}\right) }{dt}%
+D_{as}  \label{eq-force-s}
\end{equation}%
and $L_{as}$ and $D_{as}$ represent the momentum change inside the body due
to sources.

C) \emph{momentum change inside the body}. Now consider the momentum changes
inside the body due to vortices ($L_{av},D_{av}$) and sources ($%
L_{as},D_{as} $). With {\normalsize \ }$L_{a}=L_{av}+L_{as}$ and $%
D_{a}=D_{av}+D_{as}$ we may write {\normalsize \ }%
\begin{equation}
L_{a}=\underset{i}{\sum }\frac{d}{dt}\int \int_{A}\rho vdydx\text{, }D_{a}=%
\underset{i}{\sum }\frac{d}{dt}\int \int_{A}\rho udydx  \label{eq-force-la}
\end{equation}%
with the body fixed since the role due to accelerating translation and
rotation will be treated separately below.

In appendix A, we will prove that
\begin{equation}
L_{a}=0,D_{a}=0  \label{eq-force-laa}
\end{equation}

D)\emph{\ body acceleration and rotation, added mass effect}. The forces $%
(L_{add},D_{add})$ due to body acceleration and rotation have been well
studied in the past using either the kinetic energy method (cf Lamb(1932))
or the unsteady Blasius equation (cf. Wu, Yang \&Young (2012)), and have
been shown, by Wu, Yang \&Young (2012), as

\begin{equation}
\left\{
\begin{array}{c}
L_{add}=-\frac{d}{dt}\left( UA_{uu}+VA_{uv}+\Omega A_{u\Omega }\right) \\
D_{add}=-\frac{d}{dt}\left( UA_{uv}+VB_{vv}+\Omega B_{v\Omega }\right)%
\end{array}%
\right.  \label{eq-force-add}
\end{equation}%
where $A_{uu}$,$A_{uv},A_{u\Omega }$ and $B_{uu}$,$B_{uv},B_{u\Omega }$ are
added mass coefficients. The general method for computing added mass
coefficients can be found in Lamb (1932).

E) \emph{Summary}. Summing the force components defined in (\ref%
{eq-force-lift-v}),(\ref{eq-force-drag-v}), (\ref{eq-force-s}) and (\ref%
{eq-force-laa}), we obtain the lift and drag forces as
\begin{equation}
\left\{
\begin{array}{l}
L=-\rho \underset{i}{\sum }\left( V_{\infty }\Gamma _{i}-\frac{d\left(
\Gamma _{i}x_{i}\right) }{dt}\right) +\rho \underset{i}{\sum }\frac{d\left(
m_{i}y_{i}\right) }{dt}+L_{add} \\
D=-\rho \underset{i}{\sum }\frac{d\left( \Gamma _{i}y_{i}\right) }{dt}+\rho
\underset{i}{\sum }\frac{d\left( m_{i}x_{i}\right) }{dt}+D_{add}%
\end{array}%
\right.  \label{eq-force-ld}
\end{equation}%
Here each sum is performed over all the inner and outer singularities. The
relation of (\ref{eq-force-ld}) with other known theories will be discussed
in section 2.6 (Remark 2.1).

\subsection{Induced velocity method}

In (\ref{eq-force-ld}), the forces are related to the speeds of the
singularities. Now we replace the speeds of the free singularities in terms
of the induced velocities. For convenience, we {\normalsize use} the
condition (\ref{eq-gamma}) to make the term $-\rho \sum_{i}V_{\infty }\Gamma
_{i}$ disappeared in (\ref{eq-force-ld}) and rewrite (\ref{eq-force-ld}) as

{\normalsize
\begin{equation}
\left\{
\begin{array}{l}
L=\rho \underset{j,ou}{\sum }\frac{dx_{j}}{dt}\Gamma _{j}+\rho \underset{j,ou%
}{\sum }\frac{dy_{j}}{dt}m_{j}+L_{r} \\
D=-\rho \underset{j,ou}{\sum }\frac{dy_{j}}{dt}\Gamma _{j}+\rho \underset{%
j,ou}{\sum }\frac{dx_{j}}{dt}m_{j}+D_{r}%
\end{array}%
\right.  \label{eq-bb-1}
\end{equation}%
Here} $(L_{r},D_{r})$ includes all the rest terms{\normalsize \ }%
\begin{equation}
\left\{
\begin{array}{l}
L_{r}=\rho \underset{k,in}{\sum }\left( \frac{d\left( x_{k}\Gamma
_{k}\right) }{dt}+\frac{d\left( m_{k}y_{k}\right) }{dt}\right) +\rho
\underset{j,ou}{\sum }\left( x_{j}\frac{d\Gamma _{j}}{dt}+y_{j}\frac{dm_{j}}{%
dt}\right) +L_{add} \\
D_{r}=-\rho \underset{k,in}{\sum }\left( \frac{d\left( y_{k}\Gamma
_{k}\right) }{dt}-\frac{d\left( m_{k}x_{k}\right) }{dt}\right) -\rho
\underset{j,ou}{\sum }\left( y_{j}\frac{d\Gamma _{j}}{dt}-x_{j}\frac{dm_{j}}{%
dt}\right) +D_{add}%
\end{array}%
\right.  \label{eq-force-us}
\end{equation}%
{\normalsize The symbol }$\underset{j,ou}{\sum }$ means summation over all
the vortices and sources outside of the body, while $\underset{k,in}{\sum }$
means for those inside the body.

The velocity ($\frac{dx_{j}}{dt},\frac{dy_{j}}{dt}$) for any free
singularities involved in (\ref{eq-bb-1}) is due to {\normalsize \
freestream convection and induction by all the inner and outer singularities
except itself, i.e., }

\begin{equation*}
\left\{
\begin{array}{l}
\frac{dx_{j}}{dt}=V_{\infty }-\sum\limits_{k,in}\left( \Gamma
_{k}Y_{jk}-m_{k}X_{jk}\right) -\sum\limits_{l,l\neq j,ou}\left( \Gamma
_{l}Y_{jl}-m_{l}X_{jl}\right) \\
\frac{dy_{j}}{dt}=\sum\limits_{k,in}\left( \Gamma
_{k}X_{jk}+m_{k}Y_{jk}\right) +\sum\limits_{l,l\neq j,ou}\left( \Gamma
_{l}X_{jl}+m_{l}Y_{jl}\right)%
\end{array}%
\right.
\end{equation*}%
Here
\begin{equation*}
X_{jk}=\frac{x_{j}-x_{k}}{2\pi d_{lj}^{2}},Y_{jk}=\frac{y_{j}-y_{k}}{2\pi
d_{jk}^{2}}
\end{equation*}%
with $d_{jk}^{2}=(x_{j}-x_{k})^{2}+(y_{j}-y_{k})^{2}$. \ \ Inserting this
into (\ref{eq-bb-1}) to replace the factors $dx_{j}/dt$ and $dy_{j}/dt$, we
may write%
\begin{equation}
\left\{
\begin{array}{l}
L=L_{b}+L_{if}+L_{ff}+L_{t}+L_{p}+L_{add} \\
D=D_{b}+D_{if}+D_{ff}+D_{t}+D_{p}+D_{add}%
\end{array}%
\right.  \label{eq-force-in}
\end{equation}%
where the various components on the right hand sides are given and discussed
below.

a) The component ($L_{b}$,$D_{b}$) is

\begin{equation*}
L_{b}=\rho V_{\infty }\sum\limits_{j,ou}\Gamma _{j},\text{ }D_{b}=0
\end{equation*}%
{\normalsize \ Since $\Gamma _{b}=-\sum\limits_{j,ou}\Gamma _{j}$ is equal
to the circulation of the total bound vortices, we have }$L_{b}=-\rho
V_{\infty }\Gamma _{b}$ and\ $D_{b}=0$. {\normalsize \ This is just the
basic force given by the Kutta-Joukowski theorem. }

{\normalsize b) The force ($L_{if},D_{if}$) defined as%
\begin{equation*}
\left\{
\begin{array}{l}
L_{if}=-\rho \underset{j,ou}{\sum }\left( \sum\limits_{k,in}\left( \Gamma
_{k}Y_{jk}-m_{k}X_{jk}\right) \Gamma _{j}-\sum\limits_{k,in}\left( \Gamma
_{k}X_{jk}+m_{k}Y_{jk}\right) m_{j}\right) \\
D_{if}=-\rho \underset{j,ou}{\sum }\left( \sum\limits_{k,in}\left( \Gamma
_{k}X_{jk}+m_{k}Y_{jk}\right) \Gamma _{j}^{(f)}+\sum\limits_{k,in}\left(
\Gamma _{k}Y_{jk}-m_{k}X_{jk}\right) m_{j}\right)%
\end{array}%
\right.
\end{equation*}%
is due to the interaction between the inner singularities and outer
singularities. \ Putting those terms with a factor }$\Gamma _{k}$ (and
similarly $m_{k}$) together and then exchanging the order of the double sum,
as {\normalsize
\begin{equation*}
\left\{
\begin{array}{l}
L_{if}=-\rho \sum\limits_{k,in}\Gamma _{k}\underset{j,ou}{\sum }\left(
Y_{jk}\Gamma _{j}-X_{jk}m_{j}\right) +\rho \sum\limits_{k,in}m_{k}\underset{%
j,ou}{\sum }\left( X_{jk}\Gamma _{j}+Y_{jk}m_{j}\right) \\
D_{if}=-\rho \sum\limits_{k,in}\Gamma _{k}\underset{j,ou}{\sum }\left(
X_{jk}\Gamma _{j}+Y_{jk}m_{j}\right) -\rho \sum\limits_{k,in}m_{k}\underset{%
j,ou}{\sum }\left( Y_{jk}\Gamma _{j}^{(f)}-X_{jk}m_{j}\right)%
\end{array}%
\right.
\end{equation*}%
we obtain%
\begin{equation}
L_{if}=-\rho \sum\limits_{k,in}u_{k}\Gamma _{k}-\rho
\sum\limits_{k,in}m_{k}v_{k}\text{, }D_{if}=\rho
\sum\limits_{k,in}v_{k}\Gamma _{k}-\rho \sum\limits_{k,in}u_{k}m_{k}
\label{eq-bb-5d}
\end{equation}%
where }$\left( {\small u}_{k},{\small v}_{k}\right) $, defined as%
{\normalsize \ }%
\begin{equation}
{\small u}_{k}=\sum\limits_{j,ou}\left( Y_{jk}\Gamma _{j}-X_{jk}m_{j}\right)
\text{, \ }{\small v}_{k}=-\sum\limits_{j,ou}\left( X_{jk}\Gamma
_{j}+Y_{jk}m_{j}\right)  \label{v-induced}
\end{equation}%
{\normalsize \ is the fluid velocity, at the location of the inner
singularity }$(k)$, induced by all the outside singularities.{\normalsize \ }

{\normalsize c) The force ($L_{ff},D_{ff}$), defined as}%
\begin{equation*}
\left\{
\begin{array}{l}
L_{ff}=\rho \underset{j,ou}{\sum }\sum\limits_{l,l\neq j,ou}\left( \Gamma
_{l}Y_{jl}-m_{l}X_{jl}\right) \left( m_{j}-\Gamma _{j}\right) \\
D_{ff}=-\rho \underset{j,ou}{\sum }\sum\limits_{l,l\neq j,ou}\left( \Gamma
_{l}X_{jl}+m_{l}Y_{jl}\right) \left( m_{j}+\Gamma _{j}\right)%
\end{array}%
\right.
\end{equation*}%
{\normalsize is due to the mutual interaction between the free
singularities. It is obvious that the contributions to this force by each
pair of $j,l$ with $j\neq l$ }$\ $mutually cancel and thus

\begin{equation*}
L_{ff}=D_{ff}=0
\end{equation*}%
{\normalsize Hence the force due to mutual interaction between the free
singularities does not contribute to forces. }

d) {\normalsize \ The force component ($L_{t},D_{t}$) defined as }

\begin{equation*}
L_{t}=\rho \sum\limits_{k,in}\left( \frac{d\left( x_{i}\Gamma _{i}\right) }{%
dt}-\frac{d\left( y_{k}m_{k}\right) }{dt}\right) ,\text{ }D_{t}=-\rho
\sum\limits_{k,in}\left( \frac{d\left( y_{k}\Gamma _{k}\right) }{dt}+\frac{%
d\left( x_{k}m_{k}\right) }{dt}\right)
\end{equation*}%
{\normalsize is due to the motion and production of strengths of the inner
singularities. }

e) {\normalsize The force component ($L_{p},D_{p}$) defined as }
\begin{equation}
L_{p}=\rho \underset{j,ou}{\sum }\left( x_{j}\frac{d\Gamma _{j}}{dt}-y_{j}%
\frac{dm_{j}}{dt}\right) ,D_{p}=-\rho \underset{j,ou}{\sum }\left( y_{j}%
\frac{d\Gamma _{j}}{dt}+x_{j}\frac{dm_{j}}{dt}\right)  \label{eq-force-p-o}
\end{equation}%
is due to production of vortices and sources outside of the body. If there
are no vortex production, then $\frac{d\Gamma _{j}^{(f)}}{dt}=0$ for each
free vortex and $L_{p}=D_{p}=0$.

f) In the above derivation, the doublets have been grouped into vortices,
since each doublet can be represented by a vortex pair. Now we make the
force contribution due to doublets (inside the body) in an explicit form. \
{\normalsize As shown in section 2.1, each doublet of strength }$\mu _{i}$
and at position {\normalsize $(x_{i},y_{i})$ can be considered as a vortex
pair of strength \ }${\normalsize \mp }${\normalsize $\Gamma _{i}$\ at $%
(x_{i},y_{i}\pm \varepsilon )$\ with $\varepsilon \rightarrow 0$\ and $%
2\varepsilon \Gamma _{i}=\mu _{i}$. \ Apply (\ref{eq-bb-5d}) to the
corresponding vortex pairs yields, for $\varepsilon \rightarrow 0$, a force
component }$(L_{u},D_{u})$ with{\normalsize \ $L_{\mu }=\rho
\sum_{i=1}^{I_{d}}\frac{\partial u_{i}^{(\mu )}}{\partial y}2\varepsilon
\Gamma _{i}^{(\mu )}$, $D_{\mu }=-\rho \sum_{i=1}^{I_{d}}\frac{\partial
v_{i}^{(\mu )}}{\partial y}2\varepsilon \Gamma _{i}^{(\mu )}$\ , or\
\begin{equation*}
L_{\mu }=\rho \sum_{i,in}\frac{\partial u_{i}}{\partial y}\mu _{i},\text{ }%
D_{\mu }=-\rho \sum_{i,in}\frac{\partial v_{i}}{\partial y}\mu _{i}
\end{equation*}%
The various force components above will be put in compact form in section
2.4. Before doing this we would like to remark that the above analysis
appears to have given a way to interpret the physical origin of each force
component. This is not seen elsewhere according to the knowledge of the
present authors. The force due to vortex production will be further
discussed in sections 2.4-2.6. }

\subsection{Summary of the induced velocity method and multibody extension\
{\protect\normalsize \ }}

{\normalsize Now assume there are vortices (not including the doublets now),
sources and doublets inside the body, and outside of the body there are a
number of free vortices and sources. Inserting the force components defined
in items a)-e) in section 2.3 into (\ref{eq-force-in}) we obtain the force
formulas below}

{\normalsize
\begin{equation}
\left\{
\begin{array}{l}
L=-\rho V_{\infty }\Gamma _{b}+L_{ind}+L_{t}+L_{p}+L_{add} \\
D=D_{ind}+D_{t}+D_{p}+D_{add}%
\end{array}%
\right.  \label{eq-force-inducedV}
\end{equation}%
Here (}$L_{ind},D_{ind}${\normalsize ), defined as }%
\begin{equation}
\left\{
\begin{array}{l}
L_{ind}=-\sum\limits_{i,in}\rho u_{i}\Gamma _{i}-\sum\limits_{i,in}\rho
v_{i}m_{i}+\rho \sum\limits_{i,in}\frac{\partial u_{i}}{\partial y}\mu _{i}
\\
D_{ind}=\sum\limits_{i,in}\rho v_{i}\Gamma _{i}-\sum\limits_{i,in}\rho
u_{i}m_{i}-\rho \sum\limits_{i,in}\frac{\partial v_{i}}{\partial y}\mu _{i}%
\end{array}%
\right.  \label{eq-force-ind}
\end{equation}%
is due to the induced velocity effect at the inner singularities, and
{\normalsize \ $(u_{i},v_{i})$, defined by (\ref{v-induced}) and rewritten
here as}%
\begin{equation}
{\small u}_{i}=\sum\limits_{j,ou}\left( -\frac{\Gamma _{j}(y_{i}-y_{j})}{%
2\pi d_{ji}^{2}}+\frac{m_{j}(x_{i}-x_{j})}{2\pi d_{ji}^{2}}\right) \text{, }%
{\small v}_{i}{\small =}\sum\limits_{j,ou}\left( \frac{\Gamma
_{j}(x_{i}-x_{j})}{2\pi d_{ji}^{2}}+\frac{m_{j}(y_{i}-y_{j})}{2\pi d_{ji}^{2}%
}\right)  \label{eq-v-induced}
\end{equation}%
{\normalsize \ }denotes the fluid velocities induced at the location of the
inner singularities\ by all the outside singularities, \emph{not including
induction by other inner singularities}. The force component ($L_{t},D_{t}$%
), due to motion and production of inner singularities, is defined by
\begin{equation}
L_{t}=\rho \sum\limits_{i,in}\left( \frac{d\left( x_{i}\Gamma _{i}\right) }{%
dt}-\frac{d\left( y_{i}m_{i}\right) }{dt}\right) ,\text{ }D_{t}=-\rho
\sum\limits_{i,in}\left( \frac{d\left( y_{i}\Gamma _{i}\right) }{dt}+\frac{%
d\left( x_{i}m_{i}\right) }{dt}\right)  \label{eq-ld-t}
\end{equation}%
Finally the component {\normalsize ($L_{p},D_{p}$), defined by (\ref%
{eq-force-p-o}), is due to production of singularities outside of the body.
Due to Kelvin theorem of conservation of circulation, we have }$\frac{%
d\Gamma _{j}}{dt}=0$ once the vortex is produced and is moving freely%
{\normalsize . Hence it remains only those just in production. Generally,
vortices will be produced at some geometric singularities, such as the
trailing edge of a body. Moreover, we do not consider the possible case of
source production outside of the body. Then \ }%
\begin{equation}
L_{p}=\rho \underset{s}{\sum }x_{s}\frac{d\Gamma _{s}}{dt},D_{p}=-\rho
\underset{s}{\sum }y_{s}\frac{d\Gamma _{s}}{dt}  \label{eq-force-ps}
\end{equation}%
Here the summation is performed over the points on the surface of the body
where we have vortex production at rate $d\Gamma _{s}/dt$. The usefulness of
this term will become clear in for instance the treatment of Wagner problem
with vortex production.

{\normalsize The similarity and essential difference of (\ref%
{eq-force-inducedV}) comparing to the force formula of Wu, Yang \&Young
(2012) will be discussed in section 2.6 (Remark 2.3).}

Now we discuss the extension of the force formula\ {\normalsize (\ref%
{eq-force-inducedV}) }to the case of multiple bodies. This force formula has
been obtained for a single body without using pressure integration. If the
pressure $p$ is used to integrate the force, as $L=\oint\limits_{\partial
A}pdx,$ $D=-\oint\limits_{\partial A}pdy$, we of course should have the same
forces as given by {\normalsize (\ref{eq-force-inducedV})}. Now remark that
the forces in the form of {\normalsize (\ref{eq-force-inducedV}) only depend
on the induced velocities inside the body and the motion and production of
singularities inside and on the body, and that the flow pattern (such as
induced velocity and their derivatives) inside the body is the same whether
the outside singularities are free ones or bound ones (as created by another
body). We thus have the same force formula if there are outside bodies,
provided the flow induced by the outside bodies be represented by a flow
induced by equivalent singularities. } {\normalsize \ }

Consider, for the case of multiple bodies (namely body $A$, $B$, $C$, $%
\cdots $), the force formulas for the body $A$ with contours $\partial A$.
{\normalsize Then, according to the above remark, the force formula for body
}$A$ is {\normalsize
\begin{equation}
\left\{
\begin{array}{l}
L_{A}=-\rho V_{\infty }\Gamma
_{b}^{(A)}+L_{ind}^{(A)}+L_{t}^{(A)}+L_{p}^{(A)}+L_{add}^{(A)} \\
D_{A}=D_{ind}^{(A)}+D_{t}^{(A)}+D_{p}^{(A)}+D_{add}^{(A)}%
\end{array}%
\right.  \label{eq-force-md}
\end{equation}%
where }$\Gamma _{b}^{(A)}${\normalsize \ is the circulation around body }$A$%
, {\normalsize \ }%
\begin{equation*}
\left\{
\begin{array}{l}
L_{ind}^{(A)}=-\sum\limits_{i,A}\rho u_{i}\Gamma _{i}-\sum\limits_{i,A}\rho
v_{i}m_{i}+\rho \sum\limits_{i,A}\frac{\partial u_{i}}{\partial y}\mu _{i}
\\
D_{ind}^{(A)}=\sum\limits_{i,A}\rho v_{i}\Gamma _{i}-\sum\limits_{i,A}\rho
u_{i}m_{i}+\rho \sum\limits_{i,A}\frac{\partial u_{i}}{\partial x}\mu _{i}%
\end{array}%
\right.
\end{equation*}%
is the induced velocity effect, with summation performed over all
singularities inside body $A$, {\normalsize \ $(u_{i},v_{i})$ is the
velocity at }$(x_{i},y_{i})$ {\normalsize induced by all the outside
singularities and bodies. The unsteady term (}$L_{t}^{(A)},D_{t}^{(A)}$%
{\normalsize ), now defined by }%
\begin{equation*}
L_{t}^{(A)}=\rho \sum\limits_{i,A}\left( \frac{d\left( x_{i}\Gamma
_{i}\right) }{dt}-\frac{d\left( y_{i}m_{i}\right) }{dt}\right) ,\text{ }%
D_{t}^{(A)}=-\rho \sum\limits_{i,A}\left( \frac{d\left( y_{i}\Gamma
_{i}\right) }{dt}+\frac{d\left( x_{i}m_{i}\right) }{dt}\right)
\end{equation*}%
is due to the motion and production of singularities inside body $A$.
Finally,
\begin{equation*}
L_{p}^{(A)}=\rho \underset{s,A}{\sum }x_{s}\frac{d\Gamma _{s}}{dt}%
,D_{p}^{(A)}=-\rho \underset{s,A}{\sum }y_{s}\frac{d\Gamma _{s}}{dt}
\end{equation*}%
is due to the vortex production (at for instance geometric singularities) on
the surface of body $A$.

The force formulas for bodies $B$,$C$, $\cdots $ can be similarly defined.

\subsection{Force formula for distributed sources}

{\normalsize Let $\omega =\omega (x,y,t)$, $m=m(x,y,t)$\ and $\mu =\mu (x,y)$
be the continuous distributions of vortices, sources and doublets, and let $(%
\widetilde{u},\widetilde{v})=(\widetilde{u}(x,y),\widetilde{v}(x,y))$\ be
the velocity inside the body and induced by all the outside vortices,
sources and doublets, free or body generated. }With the help of Dirac delta
function we may transform the force formulasabove into integral forms.

A) \emph{Singularity velocity method}. For the force formula (\ref%
{eq-force-ld}), the integral form is

\begin{equation}
\left(
\begin{array}{l}
L=\rho \frac{d}{dt}\int \int_{D_{\infty }}\omega xdxdy+\rho \frac{d}{dt}\int
\int_{D_{\infty }}mydxdy+L_{add} \\
D=-\rho \frac{d}{dt}\int \int_{D_{\infty }}\omega ydxdy+\rho \frac{d}{dt}%
\int \int_{D_{\infty }}mxdxdy+D_{add}%
\end{array}%
\right.  \label{eq-force-d}
\end{equation}%
where $D_{\infty }$ is the region occupied by both the fluid and solid. Here
we have used $\int \int_{D_{\infty }}\omega dxdy=0$.

B) \emph{induced velocity method}. Consider only multibody problems since
the single body problem is only one such a special case. For body $A$, the
integral form for{\normalsize \ (\ref{eq-force-inducedV}) and (\ref%
{eq-force-md}) is%
\begin{equation}
\left\{
\begin{array}{l}
L_{A}=-\rho V_{\infty }\int \int_{A}\omega dxdy-\rho \int \int_{A}\left(
\widetilde{u}\omega +m\widetilde{v}-\frac{\partial \widetilde{u}}{\partial y}%
\mu \right) dxdy+L_{t}^{(A)}+L_{p}^{(A)}+L_{add}^{(A)} \\
D_{A}=\rho \int \int_{A}\left( \widetilde{v}\omega -\widetilde{u}m+\frac{%
\partial \widetilde{u}}{\partial x}\mu \right)
dxdy+D_{t}^{(A)}+D_{p}^{(A)}+D_{add}^{(A)}%
\end{array}%
\right.  \label{eq-force-dim}
\end{equation}%
where}%
\begin{equation}
L_{t}^{(A)}=\rho \frac{d}{dt}\int \int_{A}\left( x\omega -ym\right) dxdy,%
\text{ }D_{t}^{(A)}=-\rho \frac{d}{dt}\int \int_{A}\left( y\omega +xm\right)
dxdy  \label{eq-force-dim-u}
\end{equation}%
and {\normalsize \ \ }%
\begin{equation}
L_{p}^{(A)}=\rho \underset{s,A}{\sum }x_{s}\frac{d\Gamma _{s}}{dt}%
,D_{p}^{(A)}=-\rho \underset{s,A}{\sum }y_{s}\frac{d\Gamma _{s}}{dt}
\label{eq-force-dim-p}
\end{equation}%
Both (\ref{eq-force-d}) and (\ref{eq-force-dim}) will be validated against
the problem of an impulsively starting plate (section 3.3), where we have
vortex production. We should emphasize that the influence due to vortex
production inside the body is embedded in (\ref{eq-force-dim-u}).

\subsection{General remarks}

Now we provide some general remarks about the force formulas derived above,
including the connections to known theories and the new features.

\emph{Remark 2.1}. The formula (\ref{eq-force-ld}) ({\normalsize singularity
velocity method}) is rather general, including the contribution from both
the inner and outer singularities. The way to express the forces and the way
to obtain this formula are new. Its integral form (\ref{eq-force-d}) is the
same as the integral approach of Wu (1981) for inviscid two dimensional
flow, except we have an additional term due to sources. When $\Gamma _{i}$
is constant and when the body is fixed, the force formula (\ref{eq-force-ld}%
) simplifies as%
\begin{equation}
L=-\underset{i}{\sum }\rho \left( V_{\infty }-\frac{dx_{i}}{dt}\right)
\Gamma _{i},D=-\underset{i}{\sum \rho }\frac{dy_{i}}{dt}\Gamma _{i}\text{ }
\label{eq-force-p}
\end{equation}%
In the special case that the outside vortices, for instance the starting
vortices, move at the freestream speed and the internal vortices are fixed,
as in the case of steady flow, we recover the classical KJ theorem from (\ref%
{eq-force-p}) {\normalsize
\begin{equation*}
L=-\rho V_{\infty }\Gamma _{b}\text{; \ \ \ }D=0\text{ \ (KJ theory)}
\end{equation*}%
} Here $\Gamma _{b}=\underset{in}{\sum }\Gamma _{i}${\normalsize . In order
to recover the Kutta Joukowski theorem, the far away starting vortex must be
taken into account by the integral approaches of Wu (1981,p438) and Howe
(1995,p416), even for steady flow. When the reference frame is such that the
far flow field is still, an artificially built image distribution inside the
body is needed to recover the Kutta Joukowski theorem in the approach of
Saffman (1992,p48). }\ The forces expressed in the form of (\ref{eq-force-p}%
) or (\ref{eq-force-ld}) {\normalsize mean that it is instead the relative
velocity of each vortex that determines the force. For instance, for the
free vortex }$i${\normalsize , the force components are proportional to its
relative velocity (}$V_{\infty }-\frac{dx_{i}}{dt},-\frac{dy_{i}}{dt}$%
{\normalsize )}. Hence the way to express the forces here is frame
independent.

\emph{Remark 2.2}. The forces due to vortex formation, for instance in the
form of (\ref{eq-force-ps}), is strange due to the dependence on the
position $x_{s}$ and $y_{s}$. This would mean that the magnitude of the
forces depend on the choice of the reference frame. This is in fact not so
since in real problems the vortices always produce in pair due to
conservation of vorticity. This means that the production of one vortex of
circulation $\Gamma _{n}$ at $x_{n}$ (which may be some point near the
trailing edge of an airfoil) is at the consequence of the production of
another vortex of circulation $-\Gamma _{n}$ at a point $\widetilde{x}_{n}$
(which may be a point inside the body close to the trailing edge) close to $%
x_{n}$. Since the momentum due to this vortex pair is
\begin{eqnarray*}
m_{y}(t) &=&\sum_{n}\overset{X_{2}}{\underset{X_{1}}{\int }}\overset{\infty }%
{\underset{-\infty }{\int }}\rho v^{(n)}dydx \\
&=&\sum_{k}\rho \left( \frac{\Gamma _{n}}{2}\left( X_{1}+X_{2}-2x_{n}\right)
-\frac{\Gamma _{n}}{2}\left( X_{1}+X_{2}-2\widehat{x}_{n}\right) \right) \\
&=&\sum_{n}\rho \Gamma _{n}(\widehat{x}_{n}-x_{n})\text{,}
\end{eqnarray*}%
the forces due to this production, when motion is excluded, are thus
\begin{equation}
L_{p}=-\sum_{n}\rho \frac{d\Gamma _{n}}{dt}(\widehat{x}_{n}-x_{n}),D_{p}=%
\sum_{n}\rho \frac{d\Gamma _{n}}{dt}\left( \widetilde{y}_{n}-y_{n}\right)
\label{eq-force-n}
\end{equation}%
Hence the magnitude of the force due to vortex production is frame\
independent. The way to treat the vortex production outside the body in the
way of expression (\ref{eq-force-dim-p}) and that inside the body embedded
in (\ref{eq-force-dim-u}) is very convenient, see\ section 3.3 for the
Wagner problem with vortex production.

\emph{Remark 2.3 }Compared to Wu, Yang \&Young (2012), where the force
formula has been obtained directly through the unsteady Blasius equation
(though suitable only for irrotational flow so that it does not apply to the
case when vortices are produced on the surface of the body), the force
formulas {\normalsize (\ref{eq-force-inducedV})} and (\ref{eq-force-md})
based on the {\normalsize induced velocity method} are more general since
here we include the role of bound vortices and vortex production. Moreover,
the induced velocity in the force formula of Wu, Yang \&Young (2012) is due
to all the singularities (including the inner ones) while the present one
this induced velocity is only due to outside singularities (and bodies). As
remarked by them, the contributions to induced velocity effect from any pair
of interior singularities cancel out and only the contributions from the
external vortices remain. Thus both approaches yield the same force for
induced velocity effect.

\emph{Remark 2.4} The force formula (\ref{eq-force-dim}) is in fact some new
form of integral approaches, valid for multiple bodies. The past force
formulas based on integral approaches involve volume (and sometimes
boundary) integrals defined either in the entire space or in the fluid
regime, and requiring the use of auxiliary potential functions for multibody
force decomposition. The present one uses integral only defined inside the
actual body and has the advantage that it does need auxiliary potential
functions. As shown in section 3.3, it will be very convenient to be used
for airfoil problems where the internal singularity distribution can be
found through standard methods.

\section{Validation and application}

{\normalsize The force formulas given in section 2 rely on the knowledge of
the circulation of the bound vortex and its time variation, position and
speed of the vortices, sources and doublets, in discrete form or distributed
form. In this section we give several examples to demonstrate the
application of the force formulas. In section 3.1 we apply the induced
velocity method to circular cylinders for which the singularities can be
determined by the method of images. In section 3.2 we use the singularity
velocity method to study the force for Karman vortex street, which involves
an infinite number of discrete vortices and the force formula need be
specially adapted to this case. In section 3.3 we use the integral form of
both the singularity velocity method and the induced velocity method to
study the lift of a thin airfoil with vortex shedding or interacting with
another airfoil presently represented by a lumped vortex. }

\subsection{Problems of circular cylinder}

{\normalsize Here we first simplify the induced velocity method for the case
of circular cylinders, then we apply the results to study one cylinder with
a pair of outside standing vortices and a source doublet, and the problem of
two circular cylinders with given circulation. All the problems have known
solutions so that they are used to validate the present force formulas.}

\subsubsection{\protect\normalsize Simplified force formula for a circular
cylinder}

{\normalsize For each vortex $j$\ of circulation $\Gamma _{j}^{(f)}$\
outside of a circular cylinder, there is one image vortex of circulation $%
\Gamma _{j}^{(o)}=\Gamma _{j}^{(f)}$\ at the origin, and one of circulation $%
\Gamma _{j}^{(m)}=-\Gamma _{j}^{(f)}$\ at the inverse point $%
(x_{j}^{(m)},y_{j}^{(m)})$. An outside\ source doublet at $(x_{j}^{(\sigma
)},y_{j}^{(\sigma )})$\ and with strength $\sigma _{j}$, has an image
doublet of strength%
\begin{equation*}
\sigma _{j}^{(m)}=\frac{a^{2}}{f_{j}^{2}}\sigma _{j}
\end{equation*}%
at its inverse point, where $f_{j}$\ is the distance of the outside doublet
to the body center. The force formula (\ref{eq-force-inducedV}) applied here
yields the following force decomposition
\begin{equation*}
\left\{
\begin{array}{l}
L=L_{B}+L_{b}+L_{o}+L_{m}+L_{\mu }+L_{f}+L_{n} \\
D=D_{B}+D_{b}+D_{o}+D_{m}+D_{\mu }+D_{f}+D_{n}%
\end{array}%
\right.
\end{equation*}%
with}%
\begin{equation}
\left\{
\begin{array}{l}
L_{B}=-\rho V_{\infty }\Gamma _{b}\text{, }D_{B}=0\text{ \ \ \ \ \ \ \ \ \ \
\ \ \ } \\
L_{b}=-\rho \sum\limits_{i,in}\widehat{u}_{i}\Gamma _{i}^{(b)}\text{, }%
D_{b}=\rho \sum\limits_{i,in}\widehat{v}_{i}\Gamma _{i}^{(b)} \\
L_{o}=-\rho \sum\limits_{i,in}\widehat{u}_{i}\Gamma _{i}^{(o)}\text{, }%
D_{o}=\rho \sum\limits_{l=1}^{J}\widehat{v}_{i}\Gamma _{i}^{(o)} \\
L_{m}=-\rho \sum\limits_{i,in}\widehat{u}_{l}\Gamma _{l}^{(m)}\text{, }%
D_{m}=\rho \sum\limits_{i,in}\widehat{v}_{l}\Gamma _{l}^{(m)} \\
L_{\mu }=\rho \sum\limits_{i,in}\frac{\partial u_{i}}{\partial y}\mu
_{i},D_{\mu }=-\rho \sum\limits_{i,in}\frac{\partial v_{i}}{\partial y}\mu
_{i} \\
L_{\sigma }=\rho \sum\limits_{j,ou}\frac{\partial u_{j}}{\partial y}\frac{%
a^{2}\sigma _{j}}{f_{j}^{2}},D_{\sigma }=-\rho \sum\limits_{j,ou}\frac{%
\partial v_{j}}{\partial y}\frac{a^{2}\sigma _{j}}{f_{j}^{2}} \\
L_{f}=0,\ \ D_{f}=0%
\end{array}%
\right.  \label{eq-all-o}
\end{equation}%
{\normalsize Here $(L_{B},D_{B}$) is the basic bound vortex force, the
components $(L_{b},D_{b}$), $(L_{o},D_{o}$) and $(L_{m},D_{m}$) are due to
induced velocities at the locations of bound vortices, image vortices at
body center and image vortices at inverse point, respectively. \ The
component $(L_{\mu },D_{_{\mu }}$) is due to the induced velocity gradient
at the location of inner real doublets, and the component $(L_{\sigma
},D_{\sigma }$) is due to the images (at the inverse point) of the outside
doublets. \ }The velocity\ {\normalsize $\left( \widehat{u}_{i},\widehat{v}%
_{i}\right) $\ is the induced fluid velocity (induced by all the outside
vortices and source doublets) relative to the speed of the internal
singularity (}$i${\normalsize ). }Finally the component $(L_{n},D_{n})$ with
$L_{n}=-\sum_{n}\rho \frac{d\Gamma _{n}}{dt}(\widehat{x}_{n}-x_{n}),$ $%
D_{n}=\sum_{n}\rho \frac{d\Gamma _{n}}{dt}\left( \widehat{y}%
_{n}-y_{n}\right) $ is due to vortex production, {\normalsize \ $%
(x_{n},y_{n})$\ and $(\widehat{x}_{n},\widehat{y}_{n})$\ denote the position
of the vortex pair with circulation production. }

\subsubsection{\protect\normalsize Standing vortex pair behind a circular
cylinder}

{\normalsize It is well known that at moderate Reynolds numbers, the flow
around a circular cylinder involves two standing, oppositely rotating
vortices behind its cylinder. An inviscid model for this consists of two
equal and opposite point vortices, of circulation $\Gamma >0$\ and $-\Gamma
<0$, standing symmetrically behind the cylinder (Saffman 1992,p42,
Milne-Thomson 1968,p370), at the positions $\overrightarrow{x}%
_{+}=(r_{f}\cos \theta _{f},r_{f}\sin \theta _{f})$\ and $\overrightarrow{x}%
_{-}=(r_{f}\cos \theta _{f},-r_{f}\sin \theta _{f})$, respectively. }

{\normalsize The image vortices at the inverse points are respectively at $%
\overrightarrow{x}_{+}^{(m)}=(\frac{a^{2}}{r_{f}}\cos \theta _{f},\frac{a^{2}%
}{r_{f}}\sin \theta _{f})$\ and $\overrightarrow{x}_{-}^{(m)}=(\frac{a^{2}}{%
r_{f}}\cos \theta _{f},-\frac{a^{2}}{r_{f}}\sin \theta _{f})$. On the Foppl
line (see for instance Saffman 1992,p.43) defined by%
\begin{equation*}
\left( r_{f}^{2}-a^{2}\right) ^{2}=4r_{f}^{4}\sin ^{2}\left( \theta
_{f}\right) \text{ or }4r_{f}^{4}-\left( r_{f}^{2}-a^{2}\right)
^{2}=4r_{f}^{4}\cos ^{2}\left( \theta _{f}\right) ,
\end{equation*}%
the two vortices, though under the convection by stream flow and under
induction by the vortices (including images) and source doublet of strength $%
\mu =2\pi a^{2}V_{\infty }$, remain stationary if the circulation is given
by }

{\normalsize
\begin{equation*}
\Gamma _{\pm }=\mp \Gamma ,\text{ }\Gamma =4\pi V_{\infty }r_{f}\sin \left(
\theta _{f}\right) \left( 1-\frac{a^{4}}{r_{f}^{4}}\right)
\end{equation*}%
It is well known that for this case the drag vanishes, either by direct
calculation of pressure on the cylinder or by considering the vortex pair as
a source doublet at far enough distance (Saffman 1992,p43). \ Now we will
check if we recover this conclusion by the force formula in terms of the
induced velocity. The induced velocities $v_{+}^{(m)}$ and $v_{-}^{(m)}$ at
the two inverse points \ and the induced velocity $v^{(o)}$ and its
derivative $\frac{\partial v^{(o)}}{\partial y}$ at the center of the
cylinder \ are found to be }

{\normalsize
\begin{equation*}
\left(
\begin{array}{l}
v_{+}^{(m)}=\frac{a^{2}V_{\infty }\sin \left( 2\theta _{f}\right) }{r_{f}^{2}%
},v_{-}^{(m)}=-\frac{a^{2}V_{\infty }\sin \left( 2\theta _{f}\right) }{%
r_{f}^{2}} \\
v^{(o)}=0,\frac{\partial v^{(o)}}{\partial y}=-\frac{\Gamma }{\pi }\frac{%
\sin 2\theta _{f}}{r_{f}^{2}}%
\end{array}%
\right.
\end{equation*}%
}

{\normalsize The drag force due to the induced velocities at the two image
points is, due to the force relation in (\ref{eq-all-o}),
\begin{equation*}
D_{m}=\rho \Gamma _{+}^{(m)}v_{+}^{(m)}+\rho \Gamma _{-}^{(m)}v_{-}^{(m)}=%
\frac{2\rho \Gamma a^{2}V_{\infty }\sin \left( 2\theta _{f}\right) }{%
r_{f}^{2}}
\end{equation*}%
and the drag due to the doublet at the body center is, according to the
fifth relation in (\ref{eq-all-o}),
\begin{equation*}
D_{\mu }=-\rho \mu \frac{\partial v^{(o)}(0,0)}{\partial y}=-\frac{%
2a^{2}\rho V_{\infty }\Gamma \sin 2\theta _{f}}{r_{f}^{2}}
\end{equation*}%
Hence }$D_{m}${\normalsize \ cancels }$D_{\mu }$, and {\normalsize the total
drag obtained by the present method vanishes. }

\subsubsection{\protect\normalsize A doublet outside of a circular cylinder}

{\normalsize Consider a doublet of strength $\sigma $\ at $y=f$, then the
velocity induced by this doublet on the line $x=0$\ is $u^{(\mu )}(0,y)=%
\frac{\sigma }{2\pi (f-y)^{2}}$\ so that its derivatives at the center and
inverse point are
\begin{equation*}
\frac{\partial u^{(\mu )}}{\partial y}(0,0)=\frac{\sigma }{\pi f^{3}},\frac{%
\partial u^{(\sigma )}}{\partial y}(0,y_{m})=\frac{\sigma }{\pi \left(
f-a^{2}/f\right) ^{3}}
\end{equation*}%
According to the fifth and sixth relations in (\ref{eq-all-o}), if there is
a doublet of strength $\mu $\ at the center of the cylinder, there is a lift
force given by }

{\normalsize
\begin{equation*}
L_{\mu }=\rho \frac{\partial u^{(\mu )}}{\partial y}(0,0)\mu =\rho \frac{%
\sigma \mu }{\pi f^{3}}
\end{equation*}%
and there is an outside doublet, there is a lift due to the image of this
doublet
\begin{equation*}
L_{\sigma }=\rho \frac{\partial u^{(\sigma )}}{\partial y}(0,y_{m})\frac{%
\sigma a^{2}}{f^{2}}=\frac{\rho a^{2}f}{\pi \left( f^{2}-a^{2}\right) ^{3}}%
\sigma ^{2}
\end{equation*}%
The latter is the same as that given by the Blasius theorem or by Lagally
theorem (cf. Milne-Thomson,1992,p232). This is a force which points to the
doublet. }

\begin{figure}[hptb]
\centering
\includegraphics[width=0.8\textwidth]{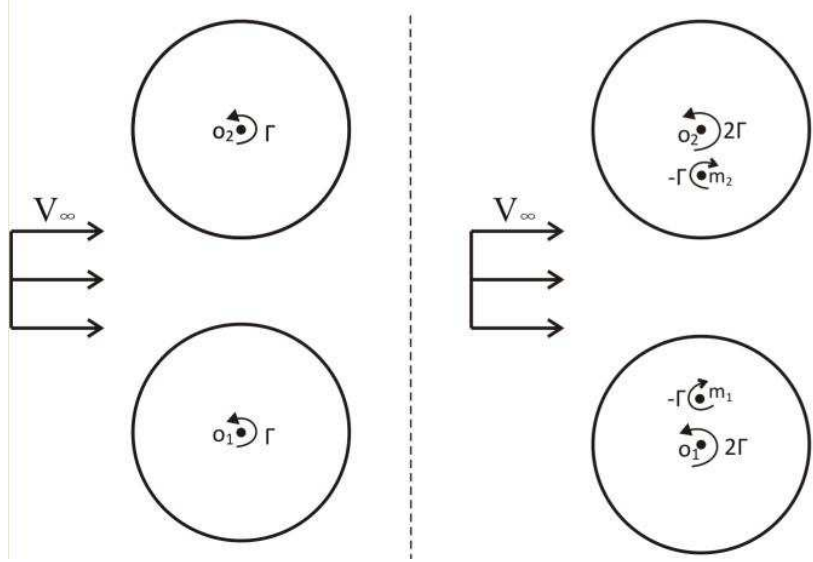}%
\caption{Staggered two cylinder problem
of Crowdy (2006). Left: staggered cylinders with bound vortices. Right:
vortex system with image vortices for the equivalent inverse point model. }%
\label{fig2}%
\end{figure}%

\subsubsection{\protect\normalsize \ Two circular cylinders with circulation}

{\normalsize Crowdy (2006) gives a general theory which permits the
calculation of lift for a finite number of staggered cylinders with bound
vortex. Here we apply (\ref{eq-all-o}) to obtain the lift for his example of
two vertically aligned cylinders (see Fig.2), both of radius $a=1/2$\ and of
given circulation $\Gamma _{1}=\Gamma _{2}=\Gamma $\ ($\Gamma =-5,-2,0$),
immersed in a uniform stream $V_{\infty }=1$\ and placed at a distance $%
h\geq 1$\ between their centers. The bound vortex associated with one
cylinder has an image pair of two counter rotating vortices in the other, at
the center and inverse point respectively. An image pair in one cylinder
also has two counter rotating image vortices in the corresponding inverse
points of the other cylinder, plus two cancelling images at the center of
the other cylinder cancel. Thus in each cylinder, there are an infinite
number of such inverse points, with counter rotating vortices of equal
strength between any two adjacent inverse points, with distance becoming
closer for newer generated images. If, for the $i$th cylinder, \ the
distance of the $k$th inverse point to the center of this cylinder is
denoted as $h_{k}^{(i)}$, then%
\begin{equation}
h_{1}^{(i)}h=a^{2}\text{, and }h_{k}^{(i)}\left( h-\frac{a^{2}}{%
h-h_{k-1}^{(i)}}\right) =a^{2}\text{ for }k>1  \label{eq-tc-1}
\end{equation}%
}

{\normalsize For an exact solution with \ (\ref{eq-all-o}), we have to work
with such infinite number of inverse points. Here we instead use an
approximate method based on the remark that the inverse points in each
cylinder are distributed in a narrow region. To see this, let $%
h_{k}^{(i)}=g^{(i)\text{ }}$\ for $k\rightarrow \infty $, then by (\ref%
{eq-tc-1}) we have $g^{(i)\text{ }}\left( h-\frac{a^{2}}{h-g^{(i)\text{ }}}%
\right) =a^{2}$, which can be solved to give }

{\normalsize
\begin{equation}
g^{(i)\text{ }}=\frac{1}{2}\left( h-\sqrt{h^{2}-4a^{2}}\right)
\label{eq-tc-2}
\end{equation}%
and one can verify that $a^{2}/h<h_{k}^{(i)}<g^{(i)\text{ }}$\ for $k>1$. It
can be further verified that the region between $y=a^{2}/h$\ and $y=g^{(i)%
\text{ }}$\ is very narrow even for $h$\ close to $1$. }

{\normalsize Hence within the framework of approximate solution we may merge
all the inverse points into an equivalent one, with a distance $h_{e}^{(i)}$%
\ to the center of the corresponding cylinder satisfying $%
a^{2}/h<h_{e}^{(i)}<g^{(i)\text{ }}$. Moreover, since these inverse points
are denser close to $y=g^{(i)\text{ }}$, we just set $h_{e}^{(i)}=a^{2}/$}$h$%
{\normalsize . With $a=\frac{1}{2}$, we have }$h_{e}^{(i)}=h_{e}=0.5^{2}/h$%
{\normalsize . \ Let }$h_{ee}${\normalsize \ be the distance between the
equivalent inverse points of the two cylinders, and }$h_{eo}$ be the
distance between the equivalent inverse point of one cylinder to the center
of the other. Then%
\begin{equation*}
h_{ee}=h-2h_{e},\text{ \ }h_{eo}=h-h_{e}
\end{equation*}

{\normalsize With the approximation of equivalent inverse point, and
considering the conservation of circulation, the vortex system in each
cylinder can be simplified in the following way. }

{\normalsize For the $1$st cylinder, there is one given bound vortex of
circulation $\Gamma _{1}^{(b)}=\Gamma _{1}$\ and one image vortex of
circulation $\Gamma _{1}^{(o)}=\Gamma _{2}$\ at the center of this cylinder,
plus one equivalent image vortex of circulation $\Gamma _{1}^{(m)}=-\Gamma
_{2}$\ at the equivalent inverse point $y=h_{e}^{(i)}$. For the $2nd$\
cylinder, this can be similarly defined. Moreover, each cylinder has a
doublet of strength $\mu _{1}=\mu _{2}=\mu =2\pi a^{2}V_{\infty }=\frac{\pi
}{2}$. There are also an infinite number of image doublets at the inverse
points. When the equivalent inverse point is applied to the doublet, the
strength of the doublet at this inverse point is }$\mu _{1}^{(e)}=\mu
_{2}^{(e)}=\mu ^{(e)}\approx \mu a^{2}/h_{eo}^{2}$. {\normalsize \ }

{\normalsize \ The velocities induced at the center and at the equivalent
inverse points of the $1st$\ and $2nd$\ cylinders by the bound and image
vortices and doublet of the $2nd$\ and $1st$\ cylinders are thus $\ $\
\begin{equation*}
\left\{
\begin{array}{c}
u_{1}^{(o)}=-u_{2}^{(o)}=\frac{2\Gamma }{2\pi h}+\frac{-\Gamma }{2\pi h_{eo}}%
+\frac{\mu }{2\pi h^{2}}+\frac{\mu ^{(e)}}{2\pi h_{eo}^{2}} \\
u_{1}^{(m)}=-u_{2}^{(m)}=\frac{2\Gamma }{2\pi h_{eo}}+\frac{-\Gamma }{2\pi
h_{ee}}+\frac{\mu }{2\pi h_{eo}^{2}}+\frac{\mu ^{(e)}}{2\pi h_{ee}^{2}}%
\end{array}%
\right.
\end{equation*}%
The derivatives of the corresponding induced velocities are
\begin{equation*}
\left\{
\begin{array}{c}
\frac{\partial u_{1}^{(\mu )}}{\partial y}=-\frac{\partial u_{2}^{(\mu )}}{%
\partial y}=\frac{2\Gamma }{2\pi h^{2}}+\frac{-\Gamma }{2\pi h_{eo}^{2}}+%
\frac{\mu }{\pi h^{3}}+\frac{\mu ^{(e)}}{\pi h_{eo}^{3}} \\
\frac{\partial u_{1}^{(\sigma )}}{\partial y}=-\frac{\partial u_{2}^{(\sigma
)}}{\partial y}=\frac{2\Gamma }{2\pi h_{eo}^{2}}+\frac{-\Gamma }{2\pi
h_{ee}^{2}}+\frac{\mu }{\pi h_{eo}^{3}}+\frac{\mu ^{(e)}}{\pi h_{ee}^{3}}%
\end{array}%
\right.
\end{equation*}%
\ According to (\ref{eq-all-o}), the force formula for the lower airfoil is }%
$L^{(1)}=L_{B}^{(1)}+L_{b}^{(1)}+L_{o}^{(1)}+L_{m}^{(1)}+L_{\mu }^{(1)}$%
{\normalsize \ with }$L_{B}^{(1)}=-\rho V_{\infty }\Gamma _{1}$, $%
L_{b}^{(1)}=-\rho u_{1}^{(o)}\Gamma _{1}$, $L_{o}^{(1)}=-\rho
u_{1}^{(o)}\Gamma _{2}^{(o)}$, $L_{m}^{(1)}=-\rho u_{1}^{(m)}\Gamma
_{2}^{(m)}$, $L_{\mu }^{(1)}=\rho \frac{\partial u_{1}^{(\mu )}}{\partial y}%
\mu _{1}$,$L_{\sigma }^{(1)}=\rho \frac{\partial u_{1}^{(\sigma )}}{\partial
y}\mu ^{(e)}$.

With {\normalsize $\rho =V_{\infty }=1$, } $\Gamma _{1}=\Gamma
_{2}^{(o)}=-\Gamma _{2}^{(m)}=\Gamma $ and $\mu _{1}=\mu =\frac{1}{2}\pi $,
we have%
\begin{equation*}
L^{(1)}=-\Gamma -2u_{1}^{(o)}\Gamma +u_{1}^{(m)}\Gamma +\frac{\partial
u_{1}^{(\mu )}}{\partial y}\mu +\frac{\partial u_{1}^{(\sigma )}}{\partial y}%
\mu ^{(e)}
\end{equation*}%
where the induced velocities $u_{1}^{(o)}$,$u_{1}^{(o)}$ and their
derivatives $\partial u_{1}^{(\mu )}/\partial y,\partial u_{1}^{(\sigma
)}/\partial y$ have been given above. {\normalsize \ \ Similarly for the
upper airfoil, }%
\begin{equation*}
L^{(2)}=-\Gamma -2u_{2}^{(o)}\Gamma +u_{2}^{(m)}\Gamma +\frac{\partial
u_{2}^{(\mu )}}{\partial y}\mu +\frac{\partial u_{2}^{(\sigma )}}{\partial y}%
\mu ^{(e)}
\end{equation*}

The results, compared to Crowdy (2006), are displayed in Fig 3. We remark
that the agreement is acceptable even when $h$ is short and despite the use
of equivalent inverse point to merge all the inverse points. The short
distance behavior, that is, there is an attraction force when the cylinders
are close, has been discussed by Crowdy. Here it is found that this is due
to the influence of the real and image doublets and the induced velocity
gradient, described by $L_{\mu }$ and $L_{\sigma }$ in (\ref{eq-all-o}).
{\normalsize {\small \ } {\small
\begin{figure}[tb]
\centering
\includegraphics[width=\textwidth]{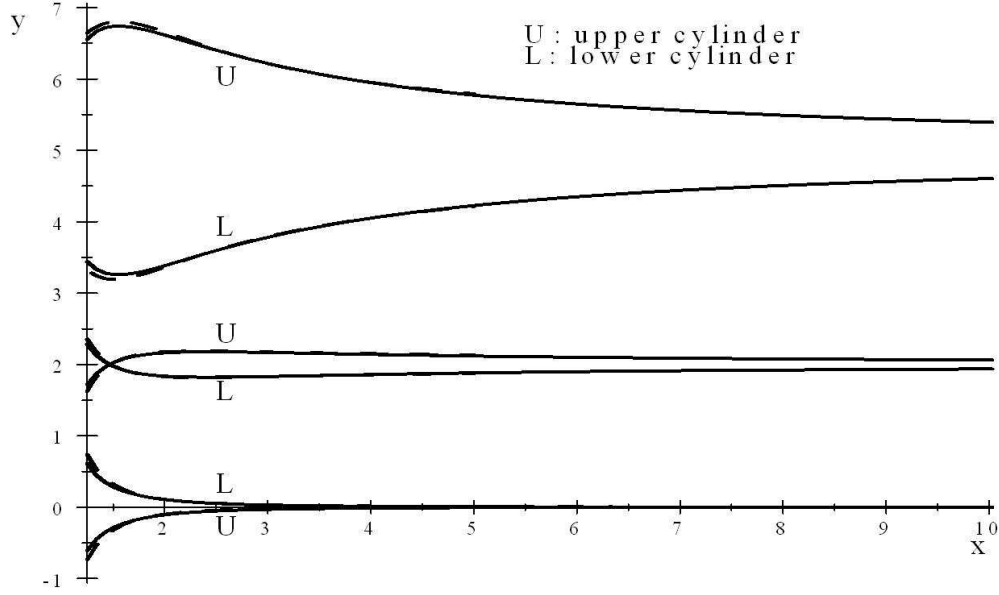}%
\caption{Comparison between the
present result (solid lines) with Crowdy (2006, Fig.4)(dashed lines), for $%
\Gamma =0$ (lower), \ $\Gamma =-2$ (middle) and $\Gamma =-5$ (upper). Dashed
lines: Crowdy. }\label{fig22}%
\end{figure}%
}}

\subsection{Karman vortex street}

{\normalsize The problem of the Karman vortex street behind a bluff body
(Fig.4) is rather special since the shape of the body is unknown. Hence we
can only use the singularity velocity method to study the drag. The Karman
vortex street is a double row of staggered and counter-rotating vortices of
strength $\Gamma >0$ and $\Gamma <0$\ and moving horizontally at speed
\begin{equation*}
V=\frac{\Gamma \pi }{a}\tanh \frac{\pi b}{a}
\end{equation*}%
Here $b$\ is the vertical separation distance between these two rows and $a$%
\ is the horizontal separation distance between adjacent two vortices in
each row (see for instance Milne-Thomson 1968,p377). The period of vortex
shedding in each row is thus
\begin{equation*}
\tau =\frac{a}{V_{\infty }-V}
\end{equation*}%
According to (\ref{eq-force-n}), the drag, averaged over $\tau $, due to the
shedding of new vortex pair, separated at a distance $\widetilde{y}%
_{n}-y_{n}=b$, is }

{\normalsize
\begin{equation*}
D_{u}=\rho \frac{1}{\tau }\int_{0}^{\tau }\frac{d\Gamma _{n}\left(
\widetilde{y}_{n}-y_{n}\right) dt}{dt}=\frac{\rho \ }{\tau }\int_{0}^{\tau
}d\left( \Gamma _{n}\left( \widetilde{y}_{n}-y_{n}\right) \right) =\frac{%
\rho b\Gamma }{\tau }
\end{equation*}%
When the expressions for }$V$ and $\tau $ are used, we obtain {\normalsize \
\begin{equation*}
D_{u}=\frac{\rho b\Gamma \left( V_{\infty }-V\right) }{a}=\frac{\rho b\Gamma
}{a}\left( V_{\infty }-\frac{\Gamma \pi }{a}\tanh \frac{\pi b}{a}\right)
\end{equation*}%
This is the well-known formula for the unsteady part of the drag, which has
been otherwise obtained by sophisticate Blasius approach (Milne-Thomosn
1968,p382), impulse approach (Saffman 1992,p 137) or integral approach
(Howe,1995,p417). The way to obtain this force using the present approach
appears to be much easier.}

\begin{figure}[ptb]
\centering
\includegraphics[width=0.6\textwidth]{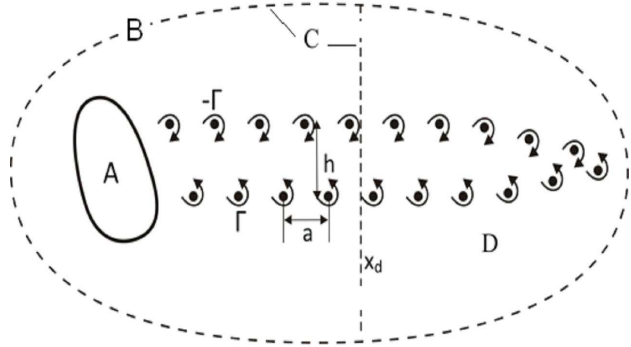}%
\caption{Control volume for the Karman
Vortex street problem. A cutline $x=x_{d}$ separates the flow regime into an
upstream part and a downstream part (D). }\label{fig33}%
\end{figure}%
{\normalsize \ }

{\normalsize Remark also that, apart from the drag component }$D_{u}$%
{\normalsize \ due to shedding of new vortices, there is also another
component }$D_{s}${\normalsize \ due to quasi steady flow formed by two rows
of periodically counter-rotating vortices spreading infinitely into the
downstream direction. No existing force formula, including the present
formula (\ref{eq-force-ld}), can be directly applied to this case. One then
needs to construct a special downstream boundary and relates this component
of drag to the momentum flux across this boundary. For details, see
Milne-Thomosn(1968,p382), (Saffman 1992,p 137) and Howe (1995,p417), where
different ways for setting the downstream boundary are used. \ Let } $%
x=x_{d} $ be such a boundary downstream of the body and intersecting the
vortex street in its uniform region. Let $(u_{k},v_{k})$ be the induced
velocity by the vortex street. Then, in Appendix B, we shall prove that, the
use of {\normalsize (\ref{eq-force-ld}) leads to}
\begin{equation}
D_{s}=\frac{1}{2}\rho \int_{x=x_{d}}\left( u_{k}^{2}-v_{k}^{2}\right) dy
\label{eq-karman}
\end{equation}%
which is exactly the same as obtained by Howe (1995, Eq.(5.11)) through an
integral approach. Howe then shows that (\ref{eq-karman}) yields\
{\normalsize \
\begin{equation*}
D_{s}=\frac{\Gamma ^{2}}{2a^{2}}\left( \frac{a}{\pi }-b\tanh \frac{\pi b}{a}%
\right)
\end{equation*}%
The total drag }$D=D_{u}+D_{s}$ {\normalsize is thus%
\begin{equation*}
D=\frac{\rho b\Gamma \left( V_{\infty }-V\right) }{a}+\frac{\Gamma ^{2}}{%
2a^{2}}\left( \frac{a}{\pi }-b\tanh \frac{\pi b}{a}\right)
\end{equation*}%
}

\subsection{\emph{Interaction of a thin airfoil with outside vortices}}

{\normalsize Consider a thin airfoil without thickness, for which the camber
line is defined by%
\begin{equation}
y=y_{a}(x)=-\alpha x+f(x),0<x<c_{A}  \label{eq-a-1-10}
\end{equation}%
Here $\alpha $\ is the angle of attack and $f(x)$\ is small with respect to $%
c_{A}$. The upstream inflow is assumed horizontal. First we need to find the
distribution of vortices. Then we apply both (\ref{eq-force-d}) (singularity
velocity method) and (\ref{eq-force-dim}) (induced velocity method) to
obtain the force formulas. The results will be validated through the problem
of Wagner with vortex shedding. Finally the interaction of the airfoil with
another one, represented by a bound vortex, is studied. }

\subsubsection{\protect\normalsize Method of solution}

{\normalsize To solve this problem we assume a distribution of vorticity $%
\gamma (x)$\ on the airfoil so that the velocity at any point on the airfoil
is%
\begin{eqnarray*}
u_{A}(x) &=&V_{\infty }-\dint\limits_{0}^{c_{A}}\frac{\gamma (\xi
)(y-y_{a}(\xi ))d\xi }{2\pi \left( (x-\xi )^{2}+(y-y_{a}(\xi ))^{2}\right) }%
+u_{v}(x)\approx V_{\infty }+u_{v}(x) \\
v_{A}(x) &=&\dint\limits_{0}^{c_{A}}\frac{\gamma (\xi )(x-\xi )d\xi }{2\pi
\left( (x-\xi )^{2}+(y-y_{a}(\xi ))^{2}\right) }+v_{v}(x)\approx
\dint\limits_{0}^{c_{A}}\frac{\gamma (\xi )d\xi }{2\pi (x-\xi )}+v_{v}(x)
\end{eqnarray*}%
where $(u_{v}(x),v_{v}(x))$\ is the velocity on the airfoil induced by
vortices outside of the airfoil and is the only correction to the standard
method for thin airfoil theory (Anderson 2010). Substituting the expressions
of }$u_{A}(x)$ and $v_{A}(x)$ {\normalsize \ into the boundary condition }

{\normalsize
\begin{equation*}
\frac{v_{A}}{u_{A}}=-\alpha +\frac{df(x)}{dx}
\end{equation*}%
we obtain the equation for $\gamma (x)$,
\begin{equation*}
\dint\limits_{0}^{c_{A}}\frac{\gamma (\xi )d\xi }{2\pi (x-\xi )}=V_{\infty
}\left( \frac{df(x)}{dx}-\alpha +g(x)\right) \text{\ }
\end{equation*}%
where the function%
\begin{equation*}
g(x)=\frac{u_{v}(x)}{V_{\infty }}\left( \frac{df(x)}{dx}-\alpha \right) -%
\frac{v_{v}(x)}{V_{\infty }}
\end{equation*}%
is due to the induction by the outside vortices. The standard thin airfoil
theory is recovered if $g(x)=0$. Let $x=\frac{c_{A}}{2}\left( 1-\cos \theta
\right) ,\xi =\frac{c_{A}}{2}\left( 1-\cos \beta \right) $\ for $0<\theta
,\beta <\pi $, then $\gamma (x)$\ can be expressed as
\begin{equation}
\gamma (\theta )=2V_{\infty }\left( A_{0}\frac{1+\cos \theta }{\sin \theta }%
+\sum_{n=1}^{n=\infty }A_{n}\sin \left( n\theta \right) \right)
\label{eq-a-1-5}
\end{equation}%
with \ $\ $\
\begin{equation}
\left\{
\begin{array}{l}
A_{0}=-\alpha +\frac{1}{\pi }\int_{0}^{\pi }\frac{df(\xi )}{d\xi }d\beta
+a_{0} \\
A_{n,n>1}=-\frac{2}{\pi }\int_{0}^{\pi }\frac{df(\xi )}{d\xi }\cos \left(
n\beta \right) d\beta +a_{n}%
\end{array}%
\right.  \label{eq-a-1-6}
\end{equation}%
where the coefficients%
\begin{equation}
\left\{
\begin{array}{l}
a_{0}=-\frac{1}{\pi V_{\infty }}\int_{0}^{\pi }\left( v_{v}(\xi )-u_{v}(\xi
)\left( \frac{df(\xi )}{d\xi }-\alpha \right) \right) d\beta \\
a_{n,n>1}=\frac{2}{\pi V_{\infty }}\int_{0}^{\pi }\left( v_{v}(\xi
)-u_{v}(\xi )\left( \frac{df(\xi )}{d\xi }-\alpha \right) \right) \cos
\left( n\beta \right) d\beta%
\end{array}%
\right.  \label{eq-a-1-7b}
\end{equation}%
are due to the outside vortices, and the remaining parts in $A_{0},A_{n}$\
are given by the classical thin airfoil theory. }

{\normalsize Remark that in the case of a thin airfoil with thickness, we
may add, in addition to the vorticity distribution $\gamma (x)$, a
distribution of source doublet $\mu (x)$\ on the camber line. Then the
boundary conditions should be defined for both the upper and lower surfaces
of the airfoil which will provide us the necessary conditions to determine
both $\gamma (x)$\ and $\mu (x)$. }

To find {\normalsize $(u_{v}(x),v_{v}(x))$ the position of each outside
vortex }$(i)$ must be determined, through solving the equations{\normalsize %
\
\begin{equation}
\left\{
\begin{array}{l}
\frac{dx_{i}}{dt}=V_{\infty }-\dint\limits_{0}^{c_{A}}\frac{\gamma (\xi
)(y_{i}-y_{a}(\xi ))d\xi }{2\pi \left( (x_{v}-\xi )^{2}+(y_{v}-y_{a}(\xi
))^{2}\right) }-\dsum\limits_{j,j\neq i}\frac{\Gamma _{j}(\xi
)(y_{i}-y_{j})d\xi }{2\pi \left( (x_{i}-x_{j})^{2}+(y_{i}-y_{j})^{2}\right) }
\\
\frac{dy_{i}}{dt}=\dint\limits_{0}^{c_{A}}\frac{\gamma (\xi )(x_{i}-\xi
)d\xi }{2\pi \left( (x-\xi )^{2}+(y-y_{a}(\xi ))^{2}\right) }%
+\dsum\limits_{j,j\neq i}\frac{\Gamma _{j}(\xi )(x_{i}-x_{j})d\xi }{2\pi
\left( (x_{i}-x_{j})^{2}+(y_{i}-y_{j})^{2}\right) }%
\end{array}%
\right.  \label{eq-1-9b}
\end{equation}%
\ }

\subsubsection{\protect\normalsize Lift force by the singularity velocity
method}

{\normalsize The lift force in the form (\ref{eq-force-d}) can be rewritten
here as%
\begin{equation*}
L=-\rho V_{\infty }\int_{0}^{c_{A}}\gamma dx+\rho \frac{d}{dt}%
\int_{0}^{c_{A}}\gamma xdx+L_{out}
\end{equation*}%
where $L_{out}$\ is due to the outside vortices. For point vortices of
circulation $\Gamma _{i}$\
\begin{equation*}
L_{out}=-\rho \sum_{i}\left( V_{\infty }-\frac{d\left( \Gamma
_{i}x_{i}\right) }{dt}\right)
\end{equation*}%
and for a distribution of outside vortices of strength $k$\ }

{\normalsize
\begin{equation}
L_{out}=-\rho V_{\infty }\int \int_{ou}kdxdy+\rho \frac{d}{dt}\int
\int_{ou}kxdxdy  \label{eq-1-6b}
\end{equation}%
\ }

{\normalsize With $\gamma (x)$\ in the form (\ref{eq-a-1-5}), we get for the
circulation of bound vortex
\begin{equation}
\Gamma _{b}=\int_{0}^{c_{A}}\gamma dx=\pi c_{A}V_{\infty }\left( A_{0}+\frac{%
1}{2}A_{1}\right)  \label{eq-1-7b}
\end{equation}%
and the moment of inner vortices }$\ $

{\normalsize
\begin{eqnarray*}
\int_{0}^{c_{A}}\gamma xdx &=&\frac{V_{\infty }c_{A}^{2}}{2}\int_{0}^{\pi
}\left( A_{0}\frac{1+\cos \theta }{\sin \theta }+\sum_{n=1}^{n=\infty
}A_{n}\sin \left( n\theta \right) \right) \left( 1-\cos \theta \right) \sin
\theta d\theta \\
&=&\frac{\pi V_{\infty }c_{A}^{2}}{4}\left( A_{0}+A_{1}-\frac{1}{2}%
A_{2}\right)
\end{eqnarray*}%
Thus the formula for the lift force is }

{\normalsize
\begin{equation}
L=-\rho V_{\infty }\Gamma _{b}+L_{int}+L_{out}  \label{eq-1-8}
\end{equation}%
where
\begin{equation}
L_{int}=\frac{\pi \rho V_{\infty }c_{A}^{2}}{4}\frac{d}{dt}\left(
A_{0}+A_{1}-\frac{A_{2}}{2}\right)  \label{eq-1-8b}
\end{equation}%
When the outside vortices are given and when $f(\xi )$\ and $\alpha $\ are
known, \ the formulas (\ref{eq-a-1-6}) and (\ref{eq-a-1-7b}) with $\xi =%
\frac{c_{A}}{2}\left( 1-\cos \beta \right) $\ yield the required
coefficients $A_{0}$, $A_{1}$\ and $A_{2}$. }

\subsubsection{\protect\normalsize Lift force by the induced velocity method}

{\normalsize If we use\ (\ref{eq-force-dim}) then
\begin{equation*}
L=-\rho V_{\infty }\int_{0}^{c_{A}}\gamma dx-\rho \int_{0}^{c_{A}}\widetilde{%
u}\gamma dx+\rho \frac{d}{dt}\int_{0}^{c_{A}}\gamma xdx+L_{p}
\end{equation*}%
where }$\int_{0}^{c_{A}}\gamma xdx=\frac{\pi V_{\infty }c_{A}^{2}}{4}\left(
A_{0}+A_{1}-\frac{1}{2}A_{2}\right) $ as above, {\normalsize \ }$L_{p}=\rho
\sum x_{s}\frac{d\Gamma _{s}}{dt}${\normalsize \ } is due to vortex
production outside of the airfoil (that due to vortex production inside the
body is included in the third term on the right hand side).

Assume that for the present problem, vortex is shed only at the trailing
edge, so that $x_{s}=c_{A},\frac{d\Gamma _{s}}{dt}=\frac{d}{dt}\int
\int_{ou}kdxdy=-\ \frac{d\Gamma _{b}}{dt}$, and
\begin{equation*}
L_{p}=-\rho c_{A}\frac{d\Gamma _{b}}{dt}
\end{equation*}%
Moreover, since the vortex sheet, {\normalsize inducing }$\widetilde{u}$, is
on the horizontal line downstream of the plate, $\int_{0}^{c_{A}}\widetilde{u%
}\gamma dx\approx 0$, and thus

\begin{equation}
L=-\rho V_{\infty }\Gamma _{b}-\rho \int_{0}^{c_{A}}\widetilde{u}\gamma dx+%
\frac{\pi \rho V_{\infty }c_{A}^{2}}{4}\frac{d}{dt}\left( A_{0}+A_{1}-\frac{%
A_{2}}{2}\right) -\rho c_{A}\frac{d\Gamma _{b}}{dt}  \label{eq-1-9}
\end{equation}

\begin{figure}[hptb]
\centering
\includegraphics[width=0.6\textwidth]{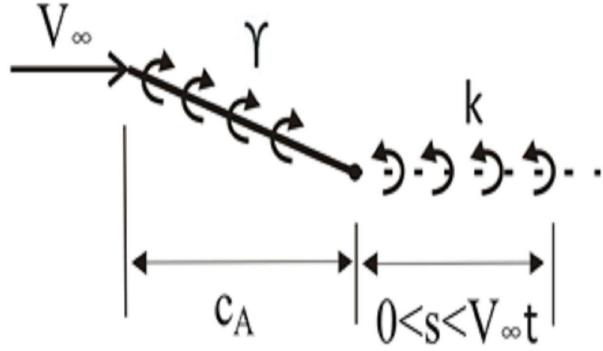}%
\caption{Impulsively started plate with
vortex shedding}\label{fig3b}%
\end{figure}%

\subsubsection{\protect\normalsize Wagner problem for impulsively starting
plate}

{\normalsize Consider a flat plate of length $c_{A}$\ at small incidence $%
\alpha $, impulsively set into motion with constant velocity $V_{\infty }$
(Fig.5). At a distance $s$\ from the trailing edge, a vortex sheet of
strength $k(s,t)$\ spreading over $0<s<V_{\infty }t$\ and satisfying }$\frac{%
Dk}{Dt}=\frac{\partial k}{\partial t}+V_{\infty }\frac{\partial k}{\partial x%
}=0$ {\normalsize under the assumption of negligible self-induced motion.
Moreover, since the total circulation is conserved, that is $\Gamma
_{b}+\int_{0}^{V_{\infty }t}k(s,t)ds=0$, we may use (\ref{eq-1-7b}) to write
the relation for determining $k(s,t)$\
\begin{equation}
\int_{0}^{V_{\infty }t}k(s,t)ds+\pi c_{A}V_{\infty }\left( A_{0}+\frac{1}{2}%
A_{1}\right) =0  \label{eq-w-2}
\end{equation}%
In Appendix C we shall prove that }

{\normalsize
\begin{equation}
A_{0}+\frac{1}{2}A_{1}=-\alpha -\frac{1}{\pi V_{\infty }c_{A}}%
\int_{0}^{V_{\infty }t}\left( 1-\frac{\sqrt{c_{A}+s}}{\sqrt{s}}\right)
k(s,t)ds  \label{eq-w-2b}
\end{equation}%
Inserting (\ref{eq-w-2b}) into (\ref{eq-w-2}), we obtain the required
equation for $k(s,t)$\ }

{\normalsize
\begin{equation}
\int_{0}^{V_{\infty }t}\frac{\sqrt{c_{A}+s}}{\sqrt{s}}k(s,t)ds=\pi
c_{A}V_{\infty }\alpha  \label{eq-w-2t}
\end{equation}%
which is exactly the same as given by the method of conformal mapping, see
for instance Saffman (1992,p111,eq (8)). }

In Appendix C we also show that{\normalsize \ \
\begin{equation*}
A_{0}+A_{1}-\frac{1}{2}A_{2}=-\alpha -\frac{1}{\pi V_{\infty }}%
\int_{0}^{V_{\infty }t}H(s)k(s,t)ds
\end{equation*}%
where $H(s)=\frac{4}{c_{A}^{2}}\left( s-\sqrt{sc_{A}+s^{2}}\right) -\frac{2}{%
c_{A}}\frac{\sqrt{c_{A}+s}}{\sqrt{s}}+\frac{4}{c_{A}}$. \ Using (\ref%
{eq-w-2t}) we may further write%
\begin{equation}
A_{0}+A_{1}-\frac{1}{2}A_{2}=\alpha -\frac{4}{\pi V_{\infty }c_{A}}%
\int_{0}^{V_{\infty }t}k(s,t)ds-\frac{4}{\pi V_{\infty }c_{A}^{2}}%
\int_{0}^{V_{\infty }t}\left( s-\sqrt{sc_{A}+s^{2}}\right) k(s,t)ds
\label{eq-w-aaa}
\end{equation}%
}$\ $

For the {\normalsize singularity velocity method},\ the component $L_{int}$
defined by{\normalsize \ (\ref{eq-1-8b}) now becomes
\begin{equation*}
L_{int}=-\rho c_{A}\frac{d}{dt}\int_{0}^{V_{\infty }t}k(s,t)ds-\rho \frac{d}{%
dt}\int_{0}^{V_{\infty }t}\left( s-\sqrt{sc_{A}+s^{2}}\right) k(s,t)ds
\end{equation*}%
when (\ref{eq-w-aaa}) is used,\ and }$L_{out}$ by{\normalsize \ (\ref%
{eq-1-6b}) \ becomes }

{\normalsize
\begin{equation*}
L_{out}=-\rho V_{\infty }\int_{0}^{V_{\infty }t}k(s,t)ds+\rho \frac{d}{dt}%
\int_{0}^{V_{\infty }t}k(s,t)\left( c_{A}+s\right) ds
\end{equation*}%
Inserting these into the force formula (\ref{eq-1-8}) and making use of $%
\Gamma _{b}+\int_{0}^{V_{\infty }t}k(s,t)ds=0$, we obtain }

{\normalsize
\begin{equation}
L=\rho \frac{d}{dt}\int_{0}^{V_{\infty }t}\sqrt{sc_{A}+s^{2}}k(s,t)ds
\label{eq-lift-w-1}
\end{equation}%
}

For the {\normalsize induced velocity method}, inserting (\ref{eq-w-aaa})
into the force formula (\ref{eq-1-9}) \ and remarking that $\int_{0}^{c_{A}}%
\widetilde{u}\gamma dx\approx 0$ since the vortex sheet, {\normalsize %
inducing }$\widetilde{u}$, is on the horizontal line downstream of the
plate, we get%
\begin{equation*}
L=-\rho V_{\infty }\Gamma _{b}-\ \left( \rho c_{A}\frac{d}{dt}%
\int_{0}^{V_{\infty }t}k(s,t)ds+\ \rho \frac{d}{dt}\int_{0}^{V_{\infty
}t}\left( s-\sqrt{sc_{A}+s^{2}}\right) k(s,t)ds\right) -\rho c_{A}\frac{%
d\Gamma _{b}}{dt}
\end{equation*}%
which, when {\normalsize $\Gamma _{b}=-\int_{0}^{V_{\infty }t}k(s,t)ds$ is
used, yields }

\begin{equation}
L=-\rho V_{\infty }\Gamma _{b}-\ \ \rho \frac{d}{dt}\int_{0}^{V_{\infty
}t}\left( s-\sqrt{sc_{A}+s^{2}}\right) k(s,t)ds  \label{eq-lift-w-2}
\end{equation}

The force formula (\ref{eq-lift-w-2}) based on the {\normalsize singularity
velocity method} and that (\ref{eq-lift-w-1}) based on the induced velocity
are in fact identical since it can be shown that
\begin{equation}
V_{\infty }\Gamma _{b}+\ \ \frac{d}{dt}\int_{0}^{V_{\infty }t}sk(s,t)ds=0
\label{eq-l-w-i}
\end{equation}%
The identity (\ref{eq-l-w-i}) will be proved below just for small time.

Now we use (\ref{eq-lift-w-2}) to study the lift for two extreme case,
{\normalsize $t\rightarrow $}${\normalsize \infty }$ and {\normalsize $%
t\rightarrow 0$. For $t\rightarrow $}${\normalsize \infty }$, Saffman
(1992,p114) shows that
\begin{equation*}
\frac{d}{dt}\int_{0}^{V_{\infty }t}\left( s-\sqrt{sc_{A}+s^{2}}\right)
k(s,t)ds=0
\end{equation*}%
Hence, by (\ref{eq-lift-w-2}),
\begin{equation*}
L=L_{\infty }=-\rho V_{\infty }\Gamma _{b}(\infty )
\end{equation*}%
where $\Gamma _{b}(\infty )=\pi V_{\infty }c_{A}\alpha $ is the steady state
circulation.

For small time {\normalsize $t\rightarrow 0$}, we may write $\left( s-\sqrt{%
sc_{A}+s^{2}}\right) ${\normalsize $\rightarrow $}$\sqrt{sc_{A}}$, thus by (%
\ref{eq-lift-w-2}),

\begin{equation*}
L\approx -\rho V_{\infty }\Gamma _{b}+\ \ \rho \frac{d}{dt}%
\int_{0}^{V_{\infty }t}\sqrt{sc_{A}}k(s,t)ds
\end{equation*}%
Furthermore, i{\normalsize t can be straightforwardly verified that the
solution for (\ref{eq-w-2t}) satisfying }$\frac{Dk}{Dt}=0${\normalsize \ in
addition is
\begin{equation*}
k(s,t)\approx \frac{V_{\infty }\alpha \sqrt{c_{A}}}{\sqrt{V_{\infty }t-s}}%
,t\rightarrow 0
\end{equation*}%
and thus
\begin{equation*}
\Gamma _{b}=-\int_{0}^{V_{\infty }t}k(s,t)ds=-2V_{\infty }\alpha \sqrt{c_{A}}%
\sqrt{V_{\infty }t}\text{, }\int_{0}^{V_{\infty }t}sk(s,t)ds\ =V_{\infty
}\alpha \sqrt{c_{A}}\frac{4\left( V_{\infty }t\right) ^{\frac{3}{2}}}{3}
\end{equation*}%
}Due to the above expressions, (\ref{eq-l-w-i}) obviously holds.

{\normalsize Now for small time, it holds that , }$\Gamma _{b}\rightarrow 0$%
{\normalsize , thus
\begin{equation*}
L\approx \rho \frac{d}{dt}\int_{0}^{V_{\infty }t}\sqrt{sc_{A}}%
k(s,t)ds\approx \rho c_{A}V_{\infty }\alpha \frac{d}{dt}\int_{0}^{V_{\infty
}t}\frac{\sqrt{s}}{\sqrt{V_{\infty }t-s}}ds
\end{equation*}%
Let $S=\sqrt{s}$, then%
\begin{equation*}
\int \frac{\sqrt{s}}{\sqrt{V_{\infty }t-s}}ds=2\int \frac{S^{2}}{\sqrt{%
V_{\infty }t-S^{2}}}dS=-S\sqrt{V_{\infty }t-S^{2}}+V_{\infty }t\arcsin \frac{%
S}{\sqrt{V_{\infty }t}}+C
\end{equation*}%
Hence $\int_{0}^{V_{\infty }t}\frac{\sqrt{s}}{\sqrt{V_{\infty }t-s}}ds=\frac{%
\pi V_{\infty }t}{2}$\ and therefore%
\begin{equation*}
L=\frac{1}{2}\pi \rho V_{\infty }^{2}c_{A}\alpha
\end{equation*}%
\ }

{\normalsize Thus we recover the well-known result that for a flat plate
impulsively generated, the initial lift on the plate is only one-half of the
final lift $L_{\infty }=\pi \rho V_{\infty }^{2}c_{A}\alpha $. \ Note also
that the initial lift for airfoils with thickness has also been studied, }%
see Chow\& Huang (1982) and Graham (1983).

\subsubsection{A bound vortex above a flat plate}

{\normalsize The lift for an airfoil with a free line vortex over the
airfoil has been studied, for instance by Saffman\&Shefield (1977) using the
method of conformal mapping. There is a condition for the vortex strength
such that the free vortex is standing. First we apply the present theory to
this case to provide another validation against known results. Then we
replace this free vortex by a bound vortex, to find the force due to
interaction between a flat plate with another airfoil represented by a
lumped vortex. \ }%
\begin{figure}[hptb]
\centering
\includegraphics[width=0.4\textwidth]{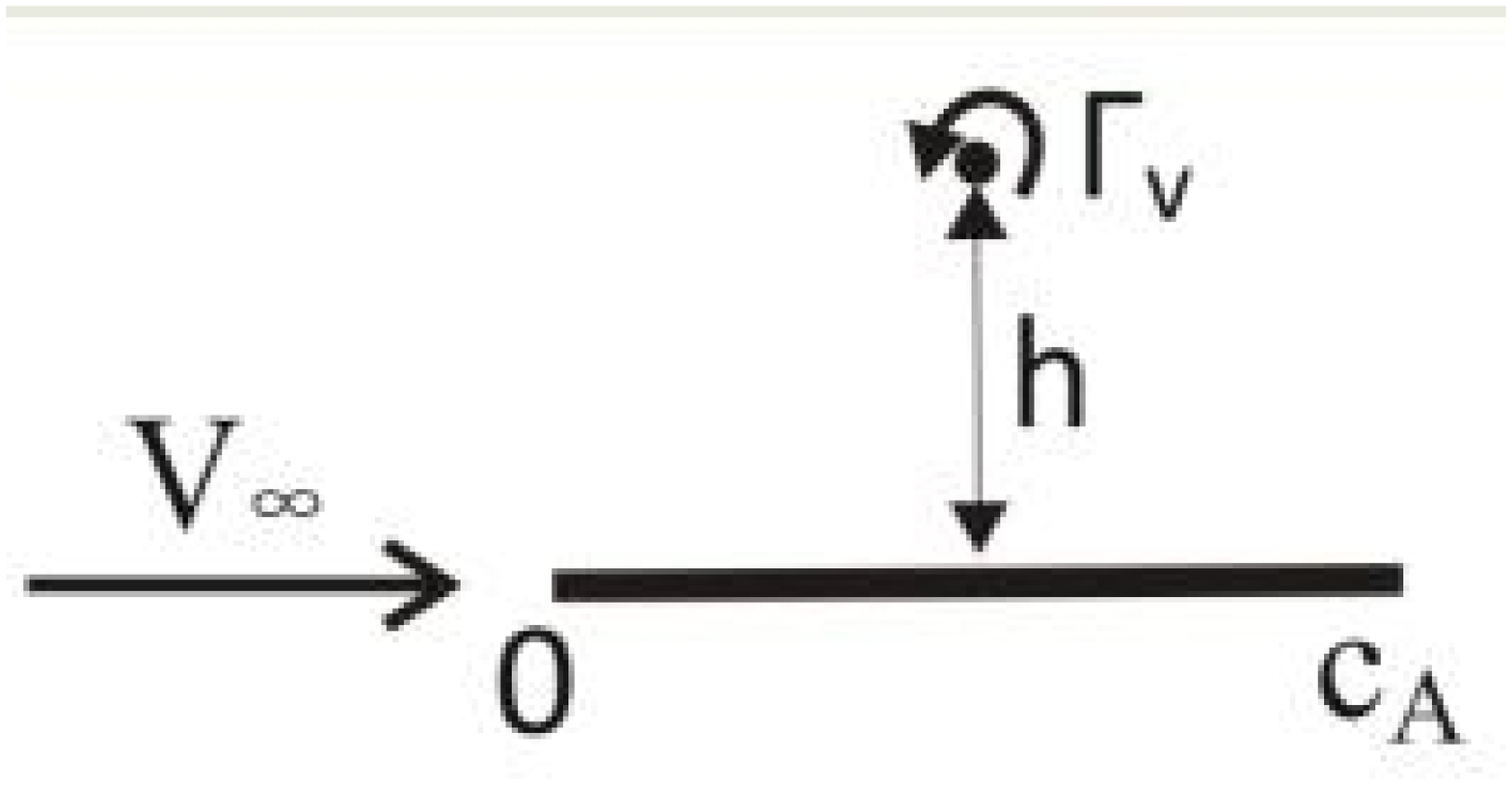}%
\label{fig5}%
\end{figure}%

Consider {\normalsize an outside vortex of strength $\Gamma _{v}>0$\ at a
distance $h$\ above the midpoint ($x_{m}=\frac{c_{A}}{2}$) of a flat plate
of length $c_{A}=2a$\ parallel to a stream of velocity $V_{\infty }$ (Fig.6)
. \ The formulas (\ref{eq-a-1-6}) and (\ref{eq-a-1-7b}) now reduce to }%
\begin{equation*}
{\normalsize A_{0}=-\frac{1}{\pi V_{\infty }}\int_{0}^{\pi }v_{v}(\xi
)d\beta }\text{, }{\normalsize \ A_{n,n>1}=\frac{2}{\pi V_{\infty }}%
\int_{0}^{\pi }v_{v}(\xi )\cos \left( n\beta \right) d\beta }
\end{equation*}%
Denote $\overline{h}=\frac{h}{\frac{1}{2}c_{A}}$. {\normalsize With $%
v_{v}(\xi )$\ given by }

{\normalsize
\begin{equation*}
v_{v}(\xi )=\frac{\Gamma _{v}}{2\pi }\frac{\xi -x_{m}}{(\xi -x_{m})^{2}+h^{2}%
}=-\frac{\Gamma _{v}}{2\pi x_{m}}\frac{\cos \beta }{\cos ^{2}\beta +%
\overline{h}^{2}},
\end{equation*}%
we get $A_{0}=A_{2}=0$ and
\begin{equation}
A_{1}=-\frac{2\Gamma _{v}}{\pi V_{\infty }c_{A}}\left( 1-\frac{\overline{h}}{%
\sqrt{1+\overline{h}^{2}}}\right) =-\frac{2\Gamma _{v}}{\pi V_{\infty }c_{A}}%
\Upsilon (\overline{h})\left( \sqrt{1+\overline{h}^{2}}+\overline{h}\right)
\label{eq-v-o-1b}
\end{equation}%
where%
\begin{equation}
\Upsilon (\overline{h})=\left( \sqrt{1+\overline{h}^{2}}+\overline{h}\right)
^{-2}\left( 1+\overline{h}^{2}\right) ^{-\frac{1}{2}}  \label{eq-laml}
\end{equation}%
Hence, with (\ref{eq-a-1-5}) and (\ref{eq-1-7b}), we obtain $\gamma (\theta
)=2V_{\infty }A_{1}\sin \theta $\ and }

{\normalsize
\begin{equation}
\Gamma _{b}=-\Upsilon (\overline{h})\left( \sqrt{1+\overline{h}^{2}}+%
\overline{h}\right) \Gamma _{v}  \label{eq-v-o-2}
\end{equation}%
This condition holds independent of whether the vortex is free or bounded.}

A) Free vortex case. \ Apply (\ref{eq-1-9b}) to the present case, and with $%
\gamma (\beta )=2V_{\infty }A_{1}\sin \beta $, we obtain the velocity of the
vortex (if it is free)
\begin{equation*}
\frac{dx_{v}}{dt}=V_{\infty }+\frac{2\Upsilon (\overline{h})\Gamma _{v}}{\pi
c_{A}}\text{, \ }\frac{dy_{v}}{dt}=0
\end{equation*}%
Set $\frac{dx_{v}}{dt}=0$, we obtain
\begin{equation}
\Gamma _{v}=-\frac{\pi V_{\infty }c_{A}}{2\Upsilon (\overline{h})}
\label{eq-v-o-3}
\end{equation}

{\normalsize The method of singularity approach (\ref{eq-1-8}) gives a lift $%
L=-\rho V_{\infty }\Gamma _{b}-\rho \left( V_{\infty }-\frac{dx_{v}}{dt}%
\right) \Gamma _{v}$, and since \ $\frac{dx_{v}}{dt}=0$, we have \ }$\ $

{\normalsize
\begin{equation}
L=\rho V_{\infty }\left( \Upsilon (\overline{h})\left( \sqrt{1+\overline{h}%
^{2}}+\overline{h}\right) -1\right) \Gamma _{v}  \label{eq-v-o-4}
\end{equation}%
}

Thus if $\Gamma _{v}$\ is given by (\ref{eq-v-o-3}) and $\Gamma _{b}$\ is
given by (\ref{eq-v-o-2}), then the outside vortex will be stationary and
the lift is given by {\normalsize (\ref{eq-v-o-4}). } {\normalsize This
result is exactly the same as given by \ the method of conformal mapping
(cf. Saffman1992,p122).}

B) Bound vortex case. This is a multibody case, though the body above the
plate is represented by a bound vortex. T{\normalsize he force formula (\ref%
{eq-1-9}) (induced velocity method) applied here gives }

{\normalsize
\begin{equation}
L=-\rho V_{\infty }\Gamma _{b}-\rho \int_{0}^{c_{A}}\widetilde{u}\gamma dx
\label{eq-v-o-5}
\end{equation}%
Here
\begin{equation*}
\widetilde{u}=u_{v}(\xi )=\frac{\Gamma _{v}}{2\pi }\frac{h}{(\xi
-x_{m})^{2}+h^{2}}=\frac{\Gamma _{v}}{2\pi x_{m}}\frac{\overline{h}}{\cos
^{2}\beta +\overline{h}^{2}}
\end{equation*}%
is the velocity in the airfoil induced by the outside bound vortex. With $%
\gamma (\beta )=2V_{\infty }A_{1}\sin \beta $\ and with $A_{1}$\ given by (%
\ref{eq-v-o-1b}), we have \
\begin{equation}
\int_{0}^{c_{A}}\widetilde{u}\gamma dx=\frac{\Gamma _{v}V_{\infty }A_{1}%
\overline{h}}{\pi }\int_{0}^{\pi }\frac{\sin \beta \sin \beta }{\cos
^{2}\beta +\overline{h}^{2}}d\beta =-\frac{2\Upsilon (\overline{h})\Gamma
_{v}^{2}}{c_{A}}  \label{eq-v-o-6}
\end{equation}%
since\ }%
\begin{equation*}
{\normalsize \dint\limits_{0}^{\pi }\frac{\sin \beta \sin \beta d\beta }{%
\cos ^{2}\beta +\overline{h}^{2}}=-\pi +\frac{\overline{h}^{2}+1}{\overline{h%
}\sqrt{1+\overline{h}^{2}}}\pi .}
\end{equation*}%
{\normalsize \ Inserting (\ref{eq-v-o-6}) and (\ref{eq-v-o-2}) into (\ref%
{eq-v-o-5}) we obtain finally
\begin{equation}
L=\rho V_{\infty }\Upsilon (\overline{h})\left( \sqrt{1+\overline{h}^{2}}+%
\overline{h}\right) \Gamma _{v}+\rho \frac{2\Upsilon (\overline{h})\Gamma
_{v}^{2}}{c_{A}}  \label{eq-v-o-7}
\end{equation}%
}

\emph{Remark 3.1}{\normalsize . The formula (\ref{eq-v-o-7}) with }$\Upsilon
(\overline{h})$ defined by{\normalsize \ (\ref{eq-laml}) gives the lift of a
horizontal flat plate with another airfoil of any given circulation }$\Gamma
_{v}$ and{\normalsize \ at a distance }$h${\normalsize \ above the middle
point of the plate, provided the additional airfoil is simplified as a
lumped vortex. }

\emph{Remark 3.2}{\normalsize . When, in addition, the circulation $\Gamma
_{v}$\ satisfies (\ref{eq-v-o-3}), then it is clear that (\ref{eq-v-o-7})
gives exactly the same force as (\ref{eq-v-o-4}). This is because that when (%
\ref{eq-v-o-3}) is satisfied, the vortex is standing for the free vortex
case so that we shall have the same force as for a bound vortex. }

Define the lift coefficient as $c_{l}=2L/\rho V_{\infty }^{2}c_{A}$ and the
normalized vortex strength as $\overline{\Gamma }_{v}=\overline{\Gamma }%
_{v}/V_{\infty }c_{A}$, we obtain from {\normalsize (\ref{eq-v-o-7}) the
lift coefficient for the flat plate}%
\begin{equation*}
c_{l}=2\left( \sqrt{1+\overline{h}^{2}}+\overline{h}+2\overline{\Gamma }%
_{v}\right) \Upsilon (\overline{h})\overline{\Gamma }_{v}
\end{equation*}%
The lift coefficient as a function of $\overline{h}$ for $\overline{\Gamma }%
_{v}=-1$ and $\overline{\Gamma }_{v}=1$ is displayed in Fig 6. When the
upper vortex has a clockwise circulation ($\overline{\Gamma }_{v}=-1$), the
lift of the airfoil, produced by interaction, is negative for $\sqrt{1+%
\overline{h}^{2}}+\overline{h}<-2\overline{\Gamma }_{v}$ or $\overline{h}<%
\frac{3}{4}$, and positive for $\overline{h}>\frac{3}{4}$. Hence there is a
distance at which the flat plate has no lift gain from the upper vortex.
When the upper vortex has anticlockwise circulation ($\overline{\Gamma }%
_{v}=1$), the lift is always positive. \
\begin{figure}[hptb]
\centering
\includegraphics[width=\textwidth]{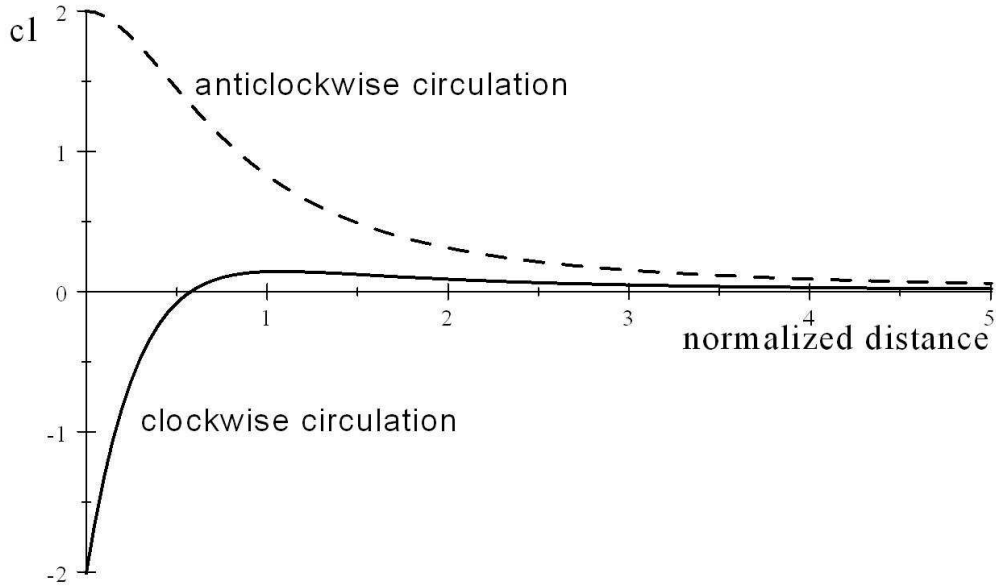}%
\caption{Lift coefficient of the flat
plate, which is interaction with a bound vortex at $\overline{h}$ above the
middle point of the plate. }\label{fig6}%
\end{figure}%

\section{Summary}

Started from a momentum balance analysis, and proceeded with inter exchange
between the singularity velocity and induced flow velocity, we have obtained
force formulas in both singularity form ( see{\normalsize \ (\ref%
{eq-force-ld}) for method of singularity velocity, and (\ref%
{eq-force-inducedV}) or (\ref{eq-force-md}) for single body or multibody
method of induced velocity) }and integral form {\normalsize \ (see (\ref%
{eq-force-d}) for method of singularity velocity and (\ref{eq-force-dim})
for multibody method of induced velocity)}, for which bound singularities,
multiple free singularities and multiple bodies can be considered. The
influence of the adjacent bodies on the actual body can be simply treated as
influence of singularities representing the adjacent bodies, in a similar
way as free singularities. Moreover, the influence on the force by vortex
production is treated in a simple and explicit way (see (\ref{eq-force-ps})
and (\ref{eq-force-dim-p}), or Remark 2.2 in section 2.6).

{\normalsize The present work is new for four reasons. \emph{First}, it
covers the work of Wu, Yang \&Young (2012) as a special case, and includes
in addition the effect of bound vortices and vortex production (see Remark
2.3 in section 2.6). \emph{Second}, the way to obtain the force formulas is
based on the interaction of various singularities, and thus is useful for
explicitly interpreting the influence due to various resources. For
instance, we have shown that the interaction between free singularities do
not contribute to forces, while the induced velocity effect is due to
interaction between free singularities and inner singularities (see section
2.3). \emph{Third}, the present result includes the situation where the
discrete singularities are replaced by a distribution of vortices and source
doublets, thus we have an integral approach which, without the use of an
auxiliary function, gives individual forces for each body in the case of
multiple bodies (see section 2.5). \emph{Last}, the present study appears to
provide a bridge between the singularity velocity approach, induced velocity
approach and some integral approaches. }

{\normalsize The present results appear to be useful for deriving analytical
force formulas even when multibody is considered. The validation study
presented in section 3 demonstrates that analytical force formulas can be
obtained even for rather complex problems. The example for an airfoil on top
of which there is another airfoil (actually represented by a bound vortex)
shows that it is possible to use the present method to derive analytical
interaction forces for multiple airfoils, and this will be for a future
study. }

The present results are actually restricted to two dimensional inviscid
flow. For viscous flow it appears that the viscous effect can be treated
separately, by simply adding an additional boundary integral (Howe 1995). \
Moreover, we did not consider the influence of body deformation, for which
well established theories exist (Landweber\&Miloh 1980, Howe 1995,
Kanso\&Oskouei 2008). Finally, the case with body acceleration and rotation
has not been tested in this paper. These problems need further studies.

Similar analysis for axisymmetric flow has been carried out for insect
flights (Wang\&Wu 2010, Wang\&Wu,2012), for which the influence of free
vortices (in the form of vortex rings, or axisymmetric Karman vortex street)
and insect body on the lift produced by the flapping wing is related to the
induced velocity. In these studies the influence of the speeds of image
vortices has been neglected, hence requiring further studies.

\appendix

\section{Momentum change inside a fixed body}

Now will show that
\begin{equation}
L_{a}\doteq \frac{d}{dt}\int \int_{A}\rho vdydx=0\text{, }D_{a}\doteq \frac{d%
}{dt}\int \int_{A}\rho udydx=0  \label{eq-ap-1a}
\end{equation}%
when the body $A$ is considered fixed.

For each vortex or source $i$ inside the body, \ we define an infinitesimal
fixed circle of radius $r_{t}\triangleq \sqrt{(x-x_{i})^{2}+(y-y_{i})^{2}}%
=\varepsilon $ whose center instantaneously coincides with the vortex. For
convenience we use here the subscript $x,y$ and $t$ denote the partial
differentials with respect to $x$,$y$ and $t$. Let $\Sigma $ be the region
bounded by the contour of the body $\partial A$ and the circumferences of
the circles $r_{i}=\varepsilon $. Below are some additional relations.

Decompose $L_{a}$ defined in (\ref{eq-ap-1a}) as \
\begin{equation*}
L_{a}=\text{\ \ }\rho \frac{d}{dt}\int \int_{\Sigma }vdydx+\rho \sum_{i,in}%
\frac{d}{dt}\int \int_{r_{i}<\varepsilon }vdydx\text{\ }
\end{equation*}%
Since $v=-\psi _{x}$ and since $\psi $ is analytical in $\Sigma $, we use
the divergence theorem to write%
\begin{equation*}
\int \int_{\Sigma }vdydx=-\int \int_{\Sigma }\psi
_{x}dydx=-\oint\limits_{\partial A}\psi
dy+\sum_{i,in}\oint\limits_{r_{i}=\varepsilon }\psi dy
\end{equation*}%
Hence%
\begin{equation*}
L_{a}=-\frac{d}{dt}\oint\limits_{\partial A}\psi dy+\sum_{i,in}\frac{d}{dt}%
\oint\limits_{r_{i}=\varepsilon }\psi dy+\rho \sum_{i,in}\frac{d}{dt}\int
\int_{r_{i}<\varepsilon }vdydx
\end{equation*}

Consider the last term on the right hand side. Across an element of length $%
r_{i}d\theta _{i}$ at $(r_{i},\theta _{i})$ on the circumference of the
circle $r_{i}<\varepsilon $, the loss of momentum due to the movement of the
vortex $(i)$ is

\begin{equation*}
dm_{i}=-v^{(i)}\left( x_{i,t}\cos \theta _{i}+y_{i,t}\sin \theta _{i}\right)
r_{i}d\theta _{i}
\end{equation*}%
Remark that $v^{(i)}=\frac{\Gamma _{i}\cos \theta _{i}}{2\pi r_{i}}$\ \ and
\begin{equation*}
\frac{d}{dt}\int \int_{r_{i}<\varepsilon }vdydx=\sum_{i,in}\int_{0}^{2\pi
}dm_{i}=-\sum_{i,in}\frac{\Gamma _{i}}{2\pi }\int_{0}^{2\pi }\cos \theta
_{i}\left( x_{i,t}\cos \theta _{i}+y_{i,t}\sin \theta _{i}\right) d\theta
_{i}
\end{equation*}%
Hence \
\begin{equation*}
\sum_{i,in}\frac{d}{dt}\int \int_{r^{(i)}<\varepsilon }vdydx=-\sum_{i,in}%
\frac{\Gamma _{i}}{2}x_{i,t}
\end{equation*}%
If sources can be similarly analyzed. \ When both vortices and sources are
present we may write%
\begin{equation*}
\sum_{i,in}\frac{d}{dt}\int \int_{r^{(i)}<\varepsilon
}vdydx=-\sum_{i,in}\left( \frac{\Gamma _{i}}{2}x_{i,t}-\frac{m_{i}}{2}%
y_{i,t}\right)
\end{equation*}%
Now consider the second term on the right hand side. First consider
vortices.
\begin{equation*}
\psi _{t}^{(i)}=-\frac{\partial }{\partial t}\left( \frac{\Gamma _{i}}{2\pi }%
\ln r_{i}\right) =-\frac{\Gamma _{i}}{2\pi r_{i}}r_{i,t}
\end{equation*}%
Let $\left( x,y\right) $\ be a fixed point on the circle. Differentiating $%
x-x_{i}=r_{i}\cos \theta _{i}$\ and $y-y_{i}=r_{i}\sin \theta _{i}$\ with
respect to time we obtain
\begin{equation*}
r_{i}\theta _{i,t}=x_{i,t}\sin \theta _{i}-y_{i,t}\cos \theta _{i}\text{, \ }%
r_{i,t}=-x_{i,t}\cos \theta _{i}-y_{i,t}\sin \theta _{i}
\end{equation*}%
Remark also that $\int_{0}^{2\pi }\cos ^{2}\theta ^{(i)}d\theta ^{(i)}=\pi $
and $\int_{0}^{2\pi }\cos \theta ^{(i)}\sin \theta ^{(i)}d\theta ^{(i)}=0$.
Thus and similarly we have\
\begin{equation*}
\oint\limits_{r_{i}=\varepsilon }\psi _{i,t}dy=-\frac{\Gamma _{i}}{2\pi }%
\int_{0}^{2\pi }r_{i,t}\cos \theta _{i}d\theta _{i}=\frac{\Gamma _{i}}{2}%
x_{i,t}
\end{equation*}%
Hence%
\begin{equation}
\sum_{i,in}\frac{d}{dt}\oint\limits_{r_{i}=\varepsilon }\psi dy=\sum_{i,in}%
\frac{\Gamma _{i}}{2}x_{i,t}  \label{eq-ss}
\end{equation}%
When sources are included, the analysis is similar and we may write%
\begin{equation*}
\sum_{i,in}\frac{d}{dt}\oint\limits_{r_{i}=\varepsilon }\psi
dy=\sum_{i,in}\left( \frac{\Gamma _{i}}{2}x_{i,t}-\frac{m_{i}}{2}%
y_{i,t}\right)
\end{equation*}%
In summary we have proved

\begin{equation*}
L_{a}=-\frac{d}{dt}\oint\limits_{\partial A}\psi dy
\end{equation*}%
Since here the body is assumed stationary so that $\psi $ is a constant
along the body, and thus $\oint\limits_{\partial A}\psi dy=0$. This means $%
L_{a}=0.$

Similarly we may prove $D_{a}=0.$

\section{Additional expression for Karman vortex street}

To prove (\ref{eq-karman}) using the present force formula {\normalsize (\ref%
{eq-force-ld})}, {\normalsize \ we need a relation between the velocities of
the inner and outer vortices. For this purpose will define a large contour\ $%
\partial C$\ enclosing the body and a part of outer vortices. Moreover, for
each outer vortex inside $\partial C$\ , we define an infinitesimal fixed
circle of radius $r_{i}=\varepsilon $\ whose center instantaneously
coincides with the vortex. Now consider the fluid region $\digamma $\
enclosed }by{\normalsize \ $\partial C$, $\partial A$\ and $\Lambda $, where
$\Lambda $\ denote the perimeters of all the fixed circles $%
r_{i}=\varepsilon $. \ We derive some integrals along \ the contours }$%
\partial {\normalsize \digamma }$.

{\normalsize Since }$\phi $ and $\psi ${\normalsize \ are analytical in $%
\digamma $, we may use the divergence theorem and the identity }$\phi
_{y}\equiv -\psi _{x}$ , $\phi _{x}\equiv \psi _{y\text{ }}$to write
\begin{eqnarray*}
\oint\limits_{\partial {\normalsize \digamma }}\left( \phi dx-\psi dy\right)
&=&-\int \int_{\digamma }\left( \phi _{y}+\psi _{x}\right) dxdy=0 \\
\oint\limits_{\partial {\normalsize \digamma }}\left( \phi dy+\psi dx\right)
&=&\int \int_{\digamma }\left( \phi _{x}-\psi _{y\text{ }}\right) dxdy=0
\end{eqnarray*}%
{\normalsize Hence}%
\begin{equation}
\left\{
\begin{array}{c}
\oint\limits_{\partial A}\left( \psi dy-\phi dx\right)
-\oint\limits_{\Lambda }\left( \phi dx-\psi dy\right)
=\oint\limits_{\partial C}\left( \psi dy-\phi dx\right) \\
\oint\limits_{\partial A}\left( \psi dx+\phi dy\right)
+\oint\limits_{\Lambda }\left( \phi dy+\psi dx\right)
=\oint\limits_{\partial C}\left( \phi dy+\psi dx\right)%
\end{array}%
\right.  \label{eq-o-1}
\end{equation}

Similarly, {\normalsize for each vortex inside $\partial A$, we define an
infinitesimal fixed circle of radius $r_{i}=\varepsilon $\ whose center
instantaneously coincides with the vortex. Now consider the region }$\Sigma $%
{\normalsize \ enclosed }by{\normalsize \ $\partial A$\ and }$\Theta $%
{\normalsize , where }$\Theta ${\normalsize \ denote the perimeters of all
the fixed circles $r_{i}=\varepsilon $ inside $\partial A$. \ }Since $\phi $
and $\psi $ are analytical inside the region $\Sigma ${\small {\normalsize ,
we can apply the }}divergence theorem to $\oint\limits_{\partial A}\left(
\psi dy-\phi dx\right) $ to write

\begin{eqnarray*}
\oint\limits_{\partial A}\left( \psi dy-\phi dx\right) -\oint\limits_{\Theta
}\left( \psi dy-\phi dx\right) &=&\int \int_{\Sigma }\left( \phi _{y}+\psi
_{x}\right) dxdy=0 \\
\oint\limits_{\partial A}\left( \psi dx+\phi dy\right) -\oint\limits_{\Theta
}\left( \psi dx+\phi dy\right) &=&\int \int_{\Sigma }\left( \phi _{x}-\psi
_{y}\right) dxdy=0
\end{eqnarray*}%
Hence%
\begin{equation*}
\oint\limits_{\partial A}\left( \psi dy-\phi dx\right) =\oint\limits_{\Theta
}\left( \psi dy-\phi dx\right) \text{, }\oint\limits_{\partial A}\left( \psi
dx+\phi dy\right) =\oint\limits_{\Theta }\left( \psi dx+\phi dy\right)
\end{equation*}%
With the above relations we may rewrite (\ref{eq-o-1}) as%
\begin{equation}
\left\{
\begin{array}{c}
\oint\limits_{\partial C}\left( \psi dy-\phi dx\right) =\oint\limits_{\Theta
}\left( \psi dy-\phi dx\right) -\oint\limits_{\Lambda }\left( \phi dx-\psi
dy\right) \\
\oint\limits_{\partial C}\left( \phi dy+\psi dx\right) =\oint\limits_{\Theta
}\left( \psi dx+\phi dy\right) +\oint\limits_{\Lambda }\left( \phi dy+\psi
dx\right)%
\end{array}%
\right.  \label{eq-k-o}
\end{equation}

As for (\ref{eq-ss}) in Appendix A, we may similarly show that

\begin{equation*}
\left\{
\begin{array}{l}
\oint\limits_{r^{(i)}=\varepsilon }\phi
_{t}dx=-\oint\limits_{r^{(i)}=\varepsilon }\psi _{t}dy=-\frac{\Gamma _{i}}{2}%
\frac{dx_{i}}{dt} \\
\oint\limits_{r^{(i)}=\varepsilon }\phi
_{t}dy=\oint\limits_{r^{(i)}=\varepsilon }\psi _{t}dx=-\frac{\Gamma _{i}}{2}%
\frac{dy_{i}}{dt}%
\end{array}%
\right.
\end{equation*}%
Inserting these expressions into (\ref{eq-k-o}), {\normalsize we obtain}

\begin{equation}
\left\{
\begin{array}{c}
\sum_{i,in}\Gamma _{i}\frac{dx_{i}}{dt}+\sum_{i,{\normalsize \digamma }%
}\Gamma _{i}\frac{dx_{i}}{dt}=\frac{d}{dt}\oint\limits_{\partial C}\left(
\psi dy-\phi dx\right) \\
-\sum_{i,in}\Gamma _{i}\frac{dy_{i}}{dt}-\sum_{i,{\normalsize \digamma }%
}\Gamma _{i}\frac{dy_{i}}{dt}=\frac{d}{dt}\oint\limits_{\partial C}\left(
\phi dy+\psi dx\right)%
\end{array}%
\right.  \label{eq-fk}
\end{equation}%
where $\sum_{i,in}$ is for vortices inside $A$ and $\sum_{i,{\normalsize %
\digamma }}$ is over vortices inside ${\normalsize \digamma }$.

The force formula (\ref{eq-force-ld}) is split here as

\begin{eqnarray*}
L &=&\rho \sum_{i,in}\Gamma _{i}\frac{dx_{i}}{dt}+\rho \underset{i,ou}{\sum }%
\Gamma _{i}\frac{dx_{i}}{dt} \\
D &=&-\rho \underset{i,in}{\sum }\Gamma _{i}\frac{dy_{i}}{dt}-\rho \underset{%
i,ou}{\sum }\Gamma _{i}\frac{dy_{i}}{dt}
\end{eqnarray*}%
which, when using (\ref{eq-fk}) to replace the first terms on the right hand
side, yields
\begin{equation}
\left\{
\begin{array}{c}
L=\rho \frac{d}{dt}\oint\limits_{\partial C}\left( \psi dy-\phi dx\right)
-\rho \underset{i,D}{\sum }\Gamma _{i}\frac{dx_{i}}{dt} \\
D=\rho \frac{d}{dt}\oint\limits_{\partial C}\left( \phi dy+\psi dx\right)
+\rho \underset{i,D}{\sum }\Gamma _{i}\frac{dy_{i}}{dt}%
\end{array}%
\right.  \label{eq-force-karman}
\end{equation}%
Here $D$ denotes the region outside of the contour $C$.

Now, we introduce a downstream boundary $x=x_{d}$ and assume that this is
the contour $\partial C$. Then with (\ref{eq-force-karman}) we may write%
\begin{equation*}
D=\rho \frac{d}{dt}\int_{x=x_{d}}\phi dy+\rho \underset{i,x_{i}>x_{d}}{\sum }%
\Gamma _{i}\frac{dy_{i}}{dt}
\end{equation*}%
{\normalsize Through defining $\omega (x,y)=\sum\limits_{j}\Gamma _{j}\delta
(x-x_{j},y-y_{j})$ (where }$\delta $ is the Dirac function{\normalsize ), we
may write}%
\begin{equation*}
\underset{i,x_{i}>x_{d}}{\sum }\Gamma _{i}\frac{dy_{i}}{dt}=\rho \frac{d}{dt}%
\int_{x>x_{c}}y\omega dxdy
\end{equation*}%
With the identity $y\omega =\nabla \cdot \left( yv,-yu\right) +u$, and the
divergence theorem so that $\int_{x>x_{d}}\nabla \cdot \left( yv,-yu\right)
dxdy=-\int_{x=x_{c}}yvdy$, we further have%
\begin{eqnarray*}
\underset{i,x_{i}>x_{d}}{\sum }\Gamma _{i}\frac{dy_{i}}{dt} &=&\rho \frac{d}{%
dt}\int_{x>x_{d}}\nabla \cdot \left( yv,-yu\right) dxdy+\rho \frac{d}{dt}%
\int_{x>x_{d}}udxdy \\
&=&-\rho \frac{d}{dt}\int_{x=x_{d}}yudy+\rho \frac{d}{dt}\int_{x>x_{d}}udxdy
\\
&=&\rho \frac{d}{dt}\int_{x=x_{d}}\phi dy+\rho \frac{d}{dt}%
\int_{x>x_{d}}udxdy
\end{eqnarray*}%
where we have used $\int_{x=x_{d}}yvdy=\int_{x=x_{d}}y\phi
_{y}dy=-\int_{x=x_{d}}\phi dy$. Hence

\begin{equation*}
D=2\rho \frac{d}{dt}\int_{x=x_{d}}\phi dy+\rho \frac{d}{dt}%
\int_{x>x_{d}}udxdy
\end{equation*}%
Using the integral form of the $y$ momentum equation for $x>x_{d}$, we have%
\begin{equation*}
\rho \frac{d}{dt}\int_{x>x_{d}}udxdy=\int_{x=x_{c}}\left( \rho u^{2}+p-\rho
u_{\infty }^{2}-p_{\infty }\right) dy
\end{equation*}%
and with the Bernoulli equation $p=-\frac{1}{2}(u^{2}+v^{2})-\phi
_{t}+p_{\infty }+\frac{1}{2}u_{\infty }^{2}$, we have%
\begin{equation*}
D=\rho \frac{d}{dt}\int_{x=x_{d}}\phi dy+\frac{1}{2}\rho
\int_{x=x_{d}}\left( u^{2}-v^{2}-u_{\infty }^{2}\right) dy
\end{equation*}

\bigskip Since the above analysis is frame independent, we may choose a
frame attached to the vortex street and therefore $\phi _{t}=0$ on the line $%
x=x_{d}$, which is assumed to intersect the vortex street in its uniform
region. Hence $D=\frac{1}{2}\rho \int_{x=x_{d}}\left( u^{2}-v^{2}-u_{\infty
}^{2}\right) dy$. Now we decompose $(u,v)$ as $u=u_{\infty }+u_{k}$, $%
v=v_{\infty }+v_{k}=v_{k}$, where $(u_{k},v_{k})$ is the induced velocity by
the vortex street, then
\begin{equation*}
D=\frac{1}{2}\rho \int_{x=x_{d}}\left( u_{k}^{2}-v_{k}^{2}\right) dy+D_{r}
\end{equation*}%
where $D_{r}=\rho u_{\infty }\int_{x=x_{d}}u_{k}dy$. It is obvious that $%
D_{r}=0$ since the contribution to $u_{k}$ by any vortex is antisymmetric
about the $y$ position of this vortex. Thus we proved (\ref{eq-karman}) in
section 3.2.

\section{Additional expression for the Wagner problem}

{\normalsize \ Remark that for the Wagner problem
\begin{equation*}
A_{0}=-\alpha -\frac{1}{\pi V_{\infty }}\int_{0}^{\pi }v_{v}(\xi )d\beta ,%
\text{ \ }A_{n,n>1}=\frac{2}{\pi V_{\infty }}\int_{0}^{\pi }v_{v}(\xi )\cos
\left( n\beta \right) d\beta
\end{equation*}%
where $v_{v}(\xi )$\ is induced by the vortex sheet and is given by
\begin{equation*}
v_{v}(\xi )=-\frac{1}{2\pi }\int_{0}^{V_{\infty }t}\frac{k(s,t)ds}{%
c_{A}+s-\xi }=-\frac{1}{2\pi }\int_{0}^{V_{\infty }t}\frac{k(s,t)}{\frac{%
c_{A}}{2}\left( \cos \beta +1\right) +s}ds
\end{equation*}%
Thus
\begin{equation*}
\left\{
\begin{array}{c}
A_{0}=-\alpha +\frac{1}{2\pi ^{2}V_{\infty }}\int_{0}^{V_{\infty
}t}\int_{0}^{\pi }\frac{d\beta }{\frac{c_{A}}{2}\left( \cos \beta +1\right)
+s}k(s,t)ds \\
A_{1}=\frac{1}{2\pi ^{2}V_{\infty }}\int_{0}^{V_{\infty }t}\int_{0}^{\pi
}\int_{0}^{\pi }\frac{-2\cos \beta d\beta }{\frac{c_{A}}{2}\left( \cos \beta
+1\right) +s}k(s,t)ds \\
A_{2}=\frac{1}{2\pi ^{2}V_{\infty }}\int_{0}^{V_{\infty }t}\int_{0}^{\pi
}\int_{0}^{\pi }\frac{-2\cos 2\beta d\beta }{\frac{c_{A}}{2}\left( \cos
\beta +1\right) +s}k(s,t)ds%
\end{array}%
\right.
\end{equation*}%
Hence }$\ $%
\begin{eqnarray*}
A_{0}+\frac{1}{2}A_{1} &=&-\alpha -\frac{1}{2\pi ^{2}V_{\infty }}%
\int_{0}^{V_{\infty }t}\int_{0}^{\pi }\frac{\left( \cos \beta -1\right)
d\beta }{\frac{c_{A}}{2}\left( \cos \beta +1\right) +s}k(s,t)d \\
A_{0}+A_{1}-\frac{1}{2}A_{2} &=&-\alpha +\frac{1}{2\pi ^{2}V_{\infty }}%
\int_{0}^{V_{\infty }t}\int_{0}^{\pi }\frac{2\left( \cos ^{2}\beta -\cos
\beta \right) d\beta }{\frac{c_{A}}{2}\left( \cos \beta +1\right) +s}k(s,t)ds
\end{eqnarray*}%
It is straightforward to show that%
\begin{equation*}
\left\{
\begin{array}{l}
A_{0}+\frac{1}{2}A_{1}=-\alpha -\frac{1}{\pi V_{\infty }c_{A}}%
\int_{0}^{V_{\infty }t}\left( 1-\frac{\sqrt{c_{A}+s}}{\sqrt{s}}\right)
{\normalsize k(s,t)}ds \\
A_{0}+A_{1}-\frac{1}{2}A_{2}=-\alpha -\frac{4}{\pi V_{\infty }c_{A}}%
\int_{0}^{V_{\infty }t}\left( {\normalsize \frac{s-\sqrt{sc_{A}+s^{2}}}{c_{A}%
}-\frac{\sqrt{c_{A}+s}}{2\sqrt{s}}+1}\right) {\normalsize k(s,t)}ds \\
\text{ \ \ \ \ \ \ \ \ \ \ \ \ \ \ }=-\alpha {\normalsize +\frac{1}{2}\frac{4%
}{\pi V_{\infty }c_{A}}\int_{0}^{V_{\infty }t}\frac{\sqrt{c_{A}+s}}{\sqrt{s}}%
k(s,t)ds} \\
\text{ \ \ \ \ \ \ \ \ \ \ \ \ \ \ \ \ }{\normalsize -}\frac{4}{\pi
V_{\infty }c_{A}}\int_{0}^{V_{\infty }t}k(s,t)ds-\frac{4}{\pi V_{\infty
}c_{A}}\int_{0}^{V_{\infty }t}{\normalsize \frac{s-\sqrt{sc_{A}+s^{2}}}{c_{A}%
}k(s,t)ds}%
\end{array}%
\right.
\end{equation*}%
{\normalsize \ by using }
\begin{equation*}
\left\{
\begin{array}{l}
\int_{0}^{\pi }\frac{d\beta }{\frac{c_{A}}{2}\left( \cos \beta +1\right) +s}=%
\frac{\pi }{\sqrt{s}\sqrt{c_{A}+s}} \\
\frac{\cos \beta -1}{\frac{c_{A}}{2}\left( \cos \beta +1\right) +s}=\frac{2}{%
c_{A}}-\frac{2\left( c_{A}+s\right) }{c_{A}\left( \frac{c_{A}}{2}\left( \cos
\beta +1\right) +s\right) } \\
\frac{\cos ^{2}\beta -\cos \beta }{\frac{c_{A}}{2}\left( \cos \beta
+1\right) +s}=\frac{2\cos \beta }{c_{A}}-\frac{4\left( c_{A}+s\right) }{%
c_{A}^{2}}+\frac{2\left( c_{A}+2s\right) \left( c_{A}+s\right) }{%
c_{A}^{2}\left( \frac{c_{A}}{2}\left( \cos \beta +1\right) +s\right) }%
\end{array}%
\right.
\end{equation*}


\begin{thebibliography}{99}
\bibitem{Anderson} Anderson J. 2010 Fundamentals of Aerodynamics,
Mcgraw-Hill Series in Aeronautical and Aerospace Engineering, McGraw-Hill
Education,New York

\bibitem{aref2007} Aref H. 2007 Point vortex dynamics: a classical
mathematics playground, Journal of Mathematical Physics. 48, 065401.

\bibitem{bw} Bai CY \& Wu ZN 2013 Generalized Kutta-Joukowski Theorem for
multi-vortices and multi-airfoil flow (lumped vortex model), Chinese Journal
of Aeronautics, accepted.

\bibitem{batchelor1967} Batchelor F.R.S. 1967 An introduction to fluid
dynamics, Cambridge University Press, Cambridge.

\bibitem{chang2008} Chang C.C., Yang S.H. \& Chu C.C. 2008 A many-body force
decomposition with applications to flow about bluff bodies, Journal of Fluid
Mechanics,600,95-104.

\bibitem{c1982} Chow C.Y. \& Huang M.K. 1982 The initial lift and drag of an
impulsively started aerofoil of finite thickness, Journal of Fluid
Mechanics. 118, 393-409.

\bibitem{dg1985} Crighton D.G. 1985 The Kutta condition in unsteady flow,
Annual Review of Fluid Mechanics, 17, 411-445.

\bibitem{crowdy2006} Crowdy D. 2006 Calculating the lift on a finite stack
of cylindrical aerofoils, Proceeding of the Royal Society A., 462, 1387-1407.

\bibitem{eams2007} Eames I, Landeryou M \& Lore JB, 2008, Inviscid coupling
between point symmetric bodies and singular distributions of vorticity,
Journal of Fluid Mechanics. 589, 33-56.

\bibitem{g1983} Graham J.M.R. 1983 The initial lift on an aerofoil in
starting flow,\ Journal of Fluid Mechanics,133, 413-425.

\bibitem{houe 1995} Howe M.S. 1995 On the force and moment on a body in an
incompressible fluid, with application to rigid bodies and bubbles at high
Reynolds numbers, Quartly Journal of Mechanics and Applied Mathematics, 48,
401-425.

\bibitem{hkcc} Hsieh C.T., Kung C.F.\&Chang C.C. 2010, Unsteady aerodynamics
of dragonfly using a simple wing-wing model from the perspective of a force
decomposition, Journal of Fluid Mechanics, 663, 233-252.

\bibitem{kanso2008} Kanso E. \& Oskouei B.G. 2008, Stability of a coupled
body--vortex system, Journal of Fluid Mechanics, 600, 77-94.

\bibitem{k2001} Katz J. \& Plotkin A. 2001 Low Speed Aerodynamics, Cambridge
University Press, Cambridge.

\bibitem{lamb} Lamb H. 1932, Hydrodynamics, Dover Publications, New York.

\bibitem{lc1989} Landweber L \& Chwang A 1989 Generalization of Taylor's
added-mass formula for two bodies, Journal of Ship Research 33, 1--9.

\bibitem{la} Landweber L \& Miloh T. 1980 Unsteady Lagally theorem for
multipoles and deformable bodies, Journal of Fluid Mechanics 96, 33-46.

\bibitem{lee1991} Lee F.J. \& Smith C.A. 1991 Effect of vortex core
distortion on blade-vortex interaction, AIAA Journal, 29,1355-1362.

\bibitem{mt1968} Milne-Thomson L.M. 1968 Theoretical Hydrodynamics,
Macmillan Education LTD, Hong Kong.

\bibitem{noca} Noca, F., Shiels, D. \& Jeon, D. 1999 A comparison of methods
for evaluating time-dependent fluid dynamic forces on bodies, using only
velocity fields and their derivatives, Journal of Fluids and Structure, 13,
551--578.

\bibitem{otbg} Oterberg D. 2010 Multi-body unsteady aerodynamics in 2D
applied to a vertical-axis wind turbine using a vortex method, Master
Thesis, Uppsala Universtity,Uppsala.

\bibitem{RAMODANOV1992} Ramodanov, SM,2002, Motion of a circular cylinder
and N point vortices in a perfect fluid, Regular and Chaotic Dynamics, 7,
291-298.

\bibitem{rr} Ragazzo, C. G. \& Tabak, E. G. 2007 On the force and torque on
systems of rigid bodies: a remark on an integral formula due to Howe.
Physics of Fluids 19, 057108.

\bibitem{saff1992} Saffman P.G. 1992 Vortex dynamics, Cambridge University
Press, New York.

\bibitem{shashi2002} Shashikanth B.N., Marsden J.E., Burdick J.W. \&Kelly
S.D. 2002 The Hamiltonian structure of a two-dimensional rigid circular
cylinder interacting dynamically with N point vortices, Physics of Fluids,
14, 1214-1227.

\bibitem{smith1996} Smith F.T. \& Timoshin S.N. 1996 Planar flows past thin
multi-blade configurations, Journal of Fluid Mechanics, 324, 355-377.

\bibitem{wangwu} Wang X.X. \& Wu Z.N. 2010 Stroke-averaged lift forces due
to vortex rings and their mutual interactions for a flapping flight model,
Journal of Fluid Mechanics, 654, 453--472.

\bibitem{ww} Wang X.X. \& Wu Z.N. 2012 Lift force reduction due to body
image of vortex for a hovering flight model, Journal of Fluid Mechanics,
709, 648-658.

\bibitem{wuct2012} Wu C.T., Yang F.L. \& Young D.L. 2012 Generalized
two-dimensional Lagally theorem with free vortices and its application to
fluid-body interaction problems, Journal of Fluid Mechanics, 698, 73--92.

\bibitem{wu1981} Wu J.C. 1981 Theory for aerodynamic force and moment in
viscous flows, AIAA Journal, 19, 432-441.

\bibitem{wlz} Wu J.C., Lu X.Y. \& Zhuang L.X. 2007 Integral force acting on
a body due to local flow structures, Journal of Fluid Mechanics, 576,
265-286.
\end{thebibliography}
\end{document}